\pdfoutput=1
\documentclass[preprint,showpacs,preprintnumbers,amsmath,amssymb,floats]{revtex4}
\usepackage{graphicx}
\usepackage{dcolumn}
\usepackage{bm}
\def\Re{\textrm{Re}}
\def\Im{\textrm{Im}}

\def\be{\begin{equation}}
\def\ee{\end{equation}}
\def\bea{\begin{eqnarray}}
\def\eea{\end{eqnarray}}

\begin{document}
\title{Considerations on the Mechanisms and \\Transition Temperatures of Superconductors}
\author{C.M. Varma}
\affiliation{Department of Physics and Astronomy, University of
California, Riverside, California 92521}

\begin{abstract}
An overview of the momentum and frequency dependence of effective electron-electron interactions which favor electronic instability to a superconducting state in the angular-momentum channel $\ell$ and the properties of the interactions which determine the magnitude of the temperature $T_c$ of the instability  is provided. Both interactions induced through exchange of phonons as well as purely electronic fluctuations of spin density, charge density or current density are considered. Special attention is paid to the role of quantum critical fluctuations including pairing due to their virtual exchange as well as de-pairing due to inelastic scattering. 

In light of the above, empirical data and theory specific to phonon induced superconductivity, superfluidity in liquid $He^3$, superconductivity in some of the heavy fermion compounds, in Cuprates, in pncitides and the valence skipping compound, is reviewed. To provide an anchor to the limits for $T_c$, the solvable case of dilute fermions with interactions varying from very weak (BCS limit) to the unitarity scattering limit and beyond (Bose-Einstein condensation of molecules) realized in experiments on cold atoms in optical traps is also discussed. 

The physical basis for the following observation is provided: The universal ratio of s-wave $T_c$ to Fermi-energy for  fermions at the unitarity limit for this last case is about 0.15,  the ratio of the maximum $T_c$ to the typical phonon frequency in phonon induced s-wave superconductivity is of the same order; the ratio of  p-wave $T_c$ to the renormalized Fermi-energy in liquid $He^3$, a very strongly correlated Fermi-liquid near its melting pressure, is only $O(10^{-3})$; in the Cuprates and the heavy-fermions where d-wave superconductivity occurs in a region governed by a special class of quantum-critical fluctuations, this ratio rises to $O(10^{-2})$. 

These discussions also suggest factors important for obtaining higher $T_c$.
Experiments and theoretical investigations are suggested to clarify the many unresolved issues. \end{abstract}
\maketitle

{\bf Table of Contents}\\

I. {\bf Introduction}. \\
   A. Scope of this Overview.\\
   B. Outline of this Overview.\\
   C. Organization and Summary of this Overview.\\
   
II. {\bf Theory of Pairing Symmetries and of $T_c$}. \\
    A. Considerations on Pairing Symmetry\\
    $~~~$ 1. Pairing Symmetry due to e-ph Interactions.\\
    $~~~$ 2. Effective Repulsion in s-wave channel in jellium \\
    $~~~$ 3. Pairing in $\ell \ne 0$ channel due to incoherent particle-hole fluctuations.\\
    $~~~$ 4. Exchange of Collective fluctuations.\\ 
    $~~~$ 5. Symmetry of Pairing Interaction from Exchange of Spin-Fluctuations \\
   B. Effect of the Frequency Dependence of Fluctuations on Tc in s-wave and higher 
       angular momentum pairing.\\ 
   C. Quantum-Criticality in Relation to in e-e Induced Pairing.\\
   D. Calculations of $T_c$ for the Hubbard Model.\\
   E. Excitonic Pairing.\\
   
III. {\bf Electron-Phonon Promoted $T_c$}.\\
   A. McMillan Expression for the Coupling constant. \\
   B. Empirical Relations in transition metal superconductivity. \\
   $~~~$1. Maximum $T_c$ and General Lessons. \\
       
IV. {\bf Superconductivity from Fermion Interactions}.\\
     A. Pairing in Dilute Fermions with varying Interactions realized in Cold Atoms in Optical Traps.\\
     B. Liquid $He^3$ : Pairing Symmetry, $T_c$ and Connection to Landau Theory.\\
     C. Empirical Results for pairing symmetry, $T_c$ and Quantum Fluctuations in Heavy 
         Fermions.\\ 
     D. Superconductivity and Quantum-criticality in the  Cuprates. \\
     E. Superconductivity in the Pnictides.\\
     F. The case of $Ba_{1-x}K_x BiO_3$.\\

 Appendix.    
\section{Introduction}

\subsection{Scope of this Overview}
There are two distinct aspects to the theory of superconductivity \cite{bcs} based on the Cooper-pair instability of the normal state of metals. The first is the theory for the kinetic energy and for the interaction vertices of electrons as a function of their momenta, spin and energy. The second is the solution of the model specified by the first aspect to understand and to calculate various properties of the superconducting state. In this overview, only the two simplest properties of the superconducting state are considered: the symmetry of the superconducting state and the transition temperature $T_c$. They are the simplest properties since they are (in most cases) provided by the linearized version of the theory. I will discuss both the well understood problem of pairing vertices in the s-wave symmetry and transition temperatures due to electron-phonon interactions (e-ph)  as well as the difficult and continuing problem of pairing due to electron-electron (e-e) interactions themselves in lower symmetries. 

The point of view taken here is that for both e-ph and e-e interactions, the second aspect, for every realized superconductor is solved by the Eliashberg extension \cite{eliashberg} of the BCS theory to include the retarded nature of the effective interactions. The validity of this theory rests on the smallness of the parameter $\lambda \omega_c/W$, where $\lambda$ is a dimensionless coupling constant, $\omega_c$ is the characteristic high frequency cut-off of the interactions and $W$ is the unrenormalized electronic bandwidth. This parameter is well known to be small enough for the e-ph problem. I will argue that for every known case of superconductivity through e-e interactions as well, this parameter is small enough for the {\it validity} of the theory if not for accurate quantitative calculations. 

This leaves the first part of the problem, which is essentially the problem about the normal state of the metal. The spin and momentum dependence of the interaction vertices can be projected onto the irreducible representations of the lattice. These {\it together} with the single-particle spectral function $A({\bf k},\omega)$  specify the possible symmetries of the superconducting state. The strength of the vertices in the different irreducible representations {\it and} the frequency dependence of the vertices are required to address quantitative questions about $T_c$ and its relationship to normal state properties. Calculation of these quantities is a forbidding prospect. Yet systematic understanding of factors determining $T_c$ appears possible with a judicious combination of empirical information, phenomenology and microscopic theory.

One may adopt contrasting attitudes towards the question of $T_c$. One attitude is that it is a foolhardy task born of naiivete. After all, we cannot even calculate the correlation energy of the normal state of a metal or of liquid $He^3$, except by complicated numerical methods which are not our purview. Calculating the frequency and momentum dependence of vertices is even harder. The other attitude is that interesting issues arise in asking the question, and relating the answers to other measurable properties, and in examining the limit of validity of the answers. The second  attitude is adopted here. This attitude is fostered by the enormous experimental and theoretical work on the e-ph problem on a wide variety of metals and the inter-relationship deduced in the various parameters determining their $T_c$. In fact physical properties of the normal state of metals, for example the electronic effects on the phonon spectra, which depend on much more detailed understanding of e-ph interactions than $T_c$, can be understood and calculated using controlled methods in agreement with experiments. The problem of e-e vertices is much harder. I will argue however that there is much to be learnt from the problem of e-ph interactions even for superconductivity due to e-e interactions. Also, there is a lot understood of the properties of 
 the normal state of liquid He$^3$ through both measurements and calculations. We do have reliable deduction of a few Landau parameters for the whole range of Liquid He$^3$ densities and systematic attempts to calculate interaction vertices by constraining them by the measured Landau parameters and relating them to transport properties and the variations of $T_c$ with density \cite{vollhardt-woelfle}.  For most metal of interest many more normal state properties are measurable in principle, and parameters fixing the essentials of the correlation functions of charge or spin or current correlations extracted. The quantitative question we shall be concerned about is the formulation of the model, the first aspect stated above, in terms of such quantities. Finally, it appears to be a fact that superconductivity of unconventional variety due to e-e interactions  is present most often only in the region of the phase diagram of metals near which a quantum-critical point \cite{cmv-phys-rep-sfl} to some other phase lies, and a good empirical case can be made that it is promoted by quantum critical fluctuations. Critical regions, once their basic physics is understood, are always easier to understand quantitatively than regions away from criticality. This is because scale-invariance of critical fluctuations guarantees that far fewer parameters are required to specify the fluctuation spectra than away from criticality and these parameters can be deduced from experiments with less ambiguity than away from critical points. 
 
On the other hand, it is important to point out that  calculations of the frequency and momentum dependent fluctuations for interacting fermions are fraught with difficulties. Especially misleading are calculations of such quantities in perturbative approximations such as the random phase or self-consistent random phase approximations or phenomenological calculations which do not respect sum-rules. The inadequacies of such calculations are evident for the case of superfluid $He^3$ and in cases where the calculations of simple models such as the Hubbard model with such methods are contrasted with those from elaborate numerical methods. The general moral is that analytical methods must be always tempered in this problem by the constraints of obtaining normal state properties in qualitative and quantitative form from the same fluctuations. 

There is a class of problems where rather precise numerical results on a model can be compared with the experiments and where at the unitarity limit of scattering universal results, independent of the details of the model are expected. This is the problem of low density of fermions on a lattice (or without a lattice) with variable inter-particle short-range attractive interaction in the s-wave channel. Definitive results obtained \cite{bec-bcs} on this problem bear comparison with experiments on cold atoms realized in optical traps \cite{coldatomExpts}. The results serve as useful anchor to the discussion of some of the more difficult problems discussed here.

  \begin{figure}
  \centering
\includegraphics[width=1.0 \textwidth]{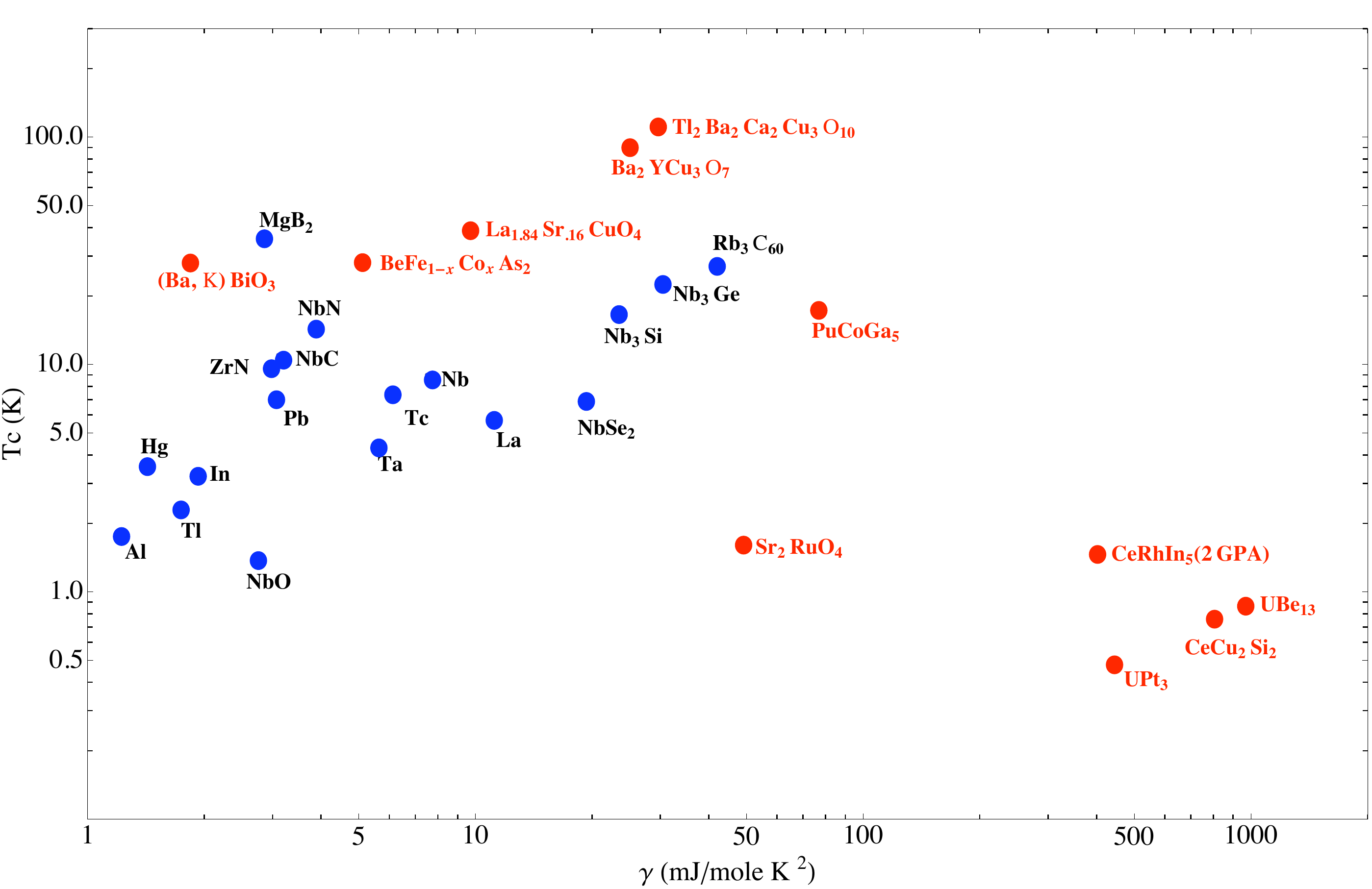}
\caption{Plot of Tc and specific heat coefficient for several classes of superconductors of interest. Black points denote those where the superconductivity is known to be promoted by electron-phonon interactions, red points denote those where it is likely that it is promoted by electron-electron interactions without the intermediary of phonons.}
\label{Tc-gamma}
\end{figure}

After this prologue, let us look at the field of action. In fig. (\ref{Tc-gamma}), measured $T_c$ for classes of superconductors is plotted on one axis and the measured specific heat coefficient $\gamma$ in the other. The black points refer to superconductors where $T_c$ is known primarily to be determined by e-ph interactions and the red points refer to the others. No direct correlation between the two quantities plotted is on the whole implied or discernible. The point of plotting $\gamma$ together with $T_c$ in the figure is simply a way of providing a parameter which is undoubtedly of importance for $T_c$ {\it within a given class of materials} as a judicious reading of this plot will reveal. In reading this figure, it should also be borne in mind that a prejudice has been made to include in the figure either those which the highest $T_c$ in their material class or those which will aid in the discussions in this paper. Included here are different classes of metals and metallic compounds including the historically first superconductor (Hg) and the old champion, the A15 compound ($Nb_3Sn,Ge$) and the new champions, ($Rb_3C_{60}$ and $MgB_2$) among those for which superconductivity is understood to be due to e-ph interactions. Among those shown in red include some cuprate superconductors, some heavy fermion superconductors, an Fe-pnictide as also the interesting case of doped $BaBiO_3$.  The crystal with the highest known $T_c \approx 164$ kelvin, $HgBa_2Ca_{m-1}Cu_{m}O_{2m+2+?}$ (Hg 1:2:m-1:m) with m=1, 2, and 3, under pressure of about 50 GPa \cite{highestTc}, is not in the plot; its specfic heat at such pressures appears not to have been measured. Superfluid  liquid $He^3$, not in the plot but which will be discussed, has a $T_c \approx 2.4 mK$ near the melting pressure  with $\gamma$ of about $50 Joule/mole K^2$. Dividing the latter by the ratio of the $He^3$ mass to the electronic mass, the equivalent electronic $\gamma$, is about $6 mJ/moleK^2$. 

The observed maximum $T_c$ from e-ph interactions, realized in metals like $Pb-Bi$ alloys, $Nb_3Sn$, $MgB_2$, etc., is of order $\theta_d/10$. Here $\theta_D$ is the Debye temperature.  For the e-e promoted cases, $T_c$, for both the heavy fermions and the Cuprates is 
$O(E_f/10^2)$ , where $E_f$ is the {\it renormalized} Fermi-energy. For $He^3$, the same ratio is about $E_f/10^{3}$. We will discuss that there are good reasons for these orders of magnitudes. I will also argue that 
$He^3$ is about as high a value of $T_c/E_f$ as one is likely to get from fermion interactions unless one is at proximity to a quantum critical point of a special kind, which is the case for both the Cuprates and the heavy-fermion compounds. 

\subsection{Organization and Summary of this overview}

I will first present in Sec. II, the minimal necessary summary of the theoretical background of aspects relevant to pairing symmetry and $T_c$. I will start with the general vertex in the pairing channel and discuss the basis for its approximation by the form generally used. I will then present the parameters that go into determining $T_c$  through Eliashberg equations for general forms of effective electron-electron interaction vertices. The parameters determining $T_c$ for the electron-phonon problem and in the general case will also be summarized. The favored "angular-momentum" state $\ell$ (more properly the irreducible representation of the point group) of pairing will be extracted from the momentum and the spin-dependence of the interaction vertices.  Also discussed here, following Kohn and Luttinger \cite{kohn-lutt}, is the instructive case of superconductivity parameters in jellium for pairing in any angular momentum. 

This, the determination of the favored channel of pairing is the easy part of the problem. The hard part is the discussion of the factors determining $T_c$. We will review that they are the spectral strength of the fluctuations exchanged, their characteristic energy and their distribution with respect to thermal energies, the coupling of these fluctuations to fermions and the density of states of the latter over the scale of the characteristic energy of the fluctuations. One of the important points emphasized here is that these factors are not independent. This is explicitly evident in the empirical evidence presented (and its derivations given in the Appendix) for the e-ph induced superconductivity. The same is expected for e-e induced superconductivity. 

I will discuss the variation of $T_c$ with the variation in the distribution in frequency of the spectral weight of the fluctuations, subject to their integral being held a constant. This includes the discussion of  the effects of inelastic scattering in depressing $T_c$ (and raising the ratio ($\Delta/T_c$)). Important differences between s-wave and finite angular momentum pairing is evident from this analysis. Using this it will be shown that the simplest class of quantum-critical fluctuations (called Gaussian fluctuations here) are bad for $T_c$.   In this class the frequency spectrum of the fluctuations is peaked near thermal frequencies. I will review that the form of critical fluctuation spectrum consistent with the normal state experiments in cuprates and the heavy-fermions and quite likely the pnictides does not suffer from this limitation.

Armed with this general knowledge, I pass on to the semi-empirical information on various classes of materials and its understanding based on the above. First superconductivity in transition metals and compounds through e-ph interaction is reviewed, in Sec. III. There are several things to be learnt here because a variety of experiments provide inter-relationships between parameters which determined $T_c$. Simple theoretical ideas have been used to understand such relationships. I argue that this understanding is useful for the e-e problems as well.

In preparation for pairing in metals due to e-e interactions, I will review the connection between vertices of Fermi-liquid theory, in particular the Landau parameters and the pairing vertex in Sec. IV. This will be accompanied by empirical information and calculations on the superfluidity of $He^3$.
I will then discuss superconductivity through other interactions. The discussion will be based both on data, empirical considerations and the theoretical guidance which is available. Superconductivity in Heavy-Fermions, Cuprates, Pnictides and the interesting case of $Ba_{1-x}K_xBiO_3$, which appears to have electronically induced $s-$ wave pairing, are successively discussed in light of the theoretical discussion. Major unsolved experimental and theoretical issues are highlighted.

What limits the $T_c/E_f$ for finite $\ell$ e-e induced pairing? There is first of all the deleterious effect on $T_c$ of the normal self-energy which is principally determined by the coupling constant $\lambda_0$ in the $s-$wave channel. Generally this is larger than the coupling constant for the pairing self-energy $\lambda_{\ell}$. This is because the important effective interactions are always short-range. Second the collective modes generally have only a fraction of the total spectra weight. Third, due to the part of the excitation spectra at thermal frequency, the role of inelastic scattering due to real as opposed to virtual scattering in pair-breaking is stronger for finite $\ell$ pairing. The gain in $T_c$ from e-e induced superconductivity of a larger characteristic energy giving a large prefactor is thus mitigated and special conditions are required for a substantive $T_c$.

The empirical results on e-ph promoted superconductors, and on classes of e-e superconductors: liquid $He^3$, Heavy-fermions, Cuprates, Pnictides and Valence-Skippers is reviewed in light of the discussions summarized above. I will explain how we understand semi-quantitatively that the maximum $T_c$ is only an order of magnitude lower than the cut-off frequency for e-ph superconductors. This is also about the upper limit obtained for the attractive Hubbard model with the same cut-off. For liquid $He^3$, the lower value $T_c/E_f$ of $O(10^{-3}$ is understood from the coupling constants obtained from the appropriate generalization of measured Landau parameters as due to the larger reduction from self-energy, the reduction of the cut-off frequency by the large renormalizations and the fact that the collective fluctuations have only a fraction of the total weight of the fluctuations given by the sum-rules. 

We gain a factor of $O(10)$ in this dimensionless ratio both for the heavy-fermion superconductors as well as the Cuprates and nearly that for the Pnictides. This is undoubtedly related to the empirical results emphasized in this overview that that high $T_c$ in Cuprates, in Heavy-Fermions, the Pnictides and the Valence-Skipper is related to proximity to a qcp of unusual variety.  As noted, quantum-criticality of the Gaussian kind is deleterious for $T_c$ because it reduces the cut-off in the fluctuations and promotes pair-breaking through inelastic scattering. The unusual nature of the fluctuations responsible for the relatively high $T_c$ of Cuprates and (in dimensionless terms) of the heavy-fermions is that the spectra though critical is distributed over a wide frequency range and that it is nearly momentum independent. The former leads to lower inelastic scattering as temperature decreases while the latter gives the smallest ratio of $\ell=0$ to finite $\ell$ coupling constants.
For Cuprates  direct evidence also shows that the cut-off frequency of this unusual spectra is only a factor of about 4 below the fermi-energy.

The derivation of the spectra for the relevant qcp for the Cuprates \cite{aji-cmv-qcf, aji-cmv-qcf-pr} relies on the discovery of an unusual competing order parameter in the Curpates. It is shown that the Action for the model can be written as a product of orthogonal topological variables, one of which has local interactions in time and logarithmic interactions in space and the other which has local interactions in space and logarithmic interaction in time. The general class of microscopic models where such properties hold for the qcf is not known; it appears empirically to contains the microscopic models that describe the Heavy-fermions and possibly the Pnictides and the valence skippers. This is an  exciting new development in the study of critical fluctuations which needs further thought and work, both experimental and theoretical.

The case of the valence skipper $Ba_xK_{1-x}BiO_3$ is discussed as it appears to be an electronically induced s-wave superconductor. Valence skippers have e-e induced pre-formed s-wave Cooper pairs in the normal state. Superconductivity consists in obtaining phase coherence of such pairs. There is some evidence, for which there is no theoretical understanding, that their fluctuation spectra is similar to that in the Cuprates. If so, not only is inelastic scattering not very deleterious, the normal self-energy from virtual processes is not as hurtful as in finite $\ell$ pairing. To my mind discovery of other valence skippers with higher electronic densities hold the best promise of further increase of $T_c$. 
 
The question is often asked, is room temperature superconductivity possible? The fact that we are only a factor of 2 away in Cuprates and we have not progressed beyond that in the past 15 years is both promising and depressing. My answer however is yes, and that the best prospects for reasons given in this paper are for electronically induced s-wave superconductors. The upper limit for $T_c/E_f$ is provided by the results on the attractive dilute Hubbard model near unitarity, which as reviewed here is $\approx 0.15$. I can only hope that, for purposes of large scale applications, this will happen in material classes which are three-dimensional, malleable in their bulk form and easy to fabricate.

\section{Theory of Pairing Symmetries and of $T_c$}

Soon after BCS theory, Eliashberg \cite{eliashberg} used the field theory methodology developed for superconductivity by Gorkov \cite{gorkov-sc} to formulate the theory of superconductivity to include the frequency dependence of the effective interactions through exchange of phonons. This theory is valid if $(\lambda \omega_c/W) << 1$. The physical content of this limit is the Migdal theorem which proves that the vertex corrections in the theory of electron-phonon interactions are of $O(\lambda \omega_c/W)$ compared to the leading dimensionless vertex $\lambda$.  

The experimental proof that the superconductivity in metals such as $Pb, Sn, Nb$, etc. is induced by electron-phonon interaction rests on analysis of tunneling spectrum \cite{mcmillan-rowell, 
schrieffer-scalapino} in these metals using the Eliashberg theory and the measurement of the spectrum of phonons by neutron scattering. The theory also provides experimental proofs of its limit of validity.

With one important modification, which does not affect the linearized Eliashberg theory which is enough to determine $T_c$,  the theory can be used for pairing in any symmetry of degenerate fermions due to exchange of any kind of fluctuations, provided they can be regarded as bosons, and provided the Migdal condition is satisfied, where $\omega_c$ is the upper cutoff of the spectrum of bosons. 

Excellent reviews of the technical aspects of the Eliashberg theory have been written \cite{Scalapino}. To fulfill the limited goals of this review, I need to present only the dependence of some of the results of the theory on the parameters of the starting model.
The model specifies a spectrum of fermions, a spectrum of bosons, and the interactions between the fermions and the bosons:
\be
H_{f} = \sum_{{\bf k}, \sigma} \epsilon_{\bf k} c_{{\bf k} \sigma}^+c_{{\bf k} \sigma}
\ee

The transition temperature is the temperature at which the particle-particle scattering vertex with total momentum $0$ and energy $0$ diverges.  The exact vertex 
$\Gamma_{S}({k, k+q})$ for fermions with total spin $S = 0,1$, scattering from ${\bf k,-k}$ to ${\bf k+q, -k-q}$, with energy $\omega$ to $\omega + \nu$ follows the Bethe-Salpeter equation through which it is related to the  irreducible particle-particle vertex $I_{S}({k, k+q})$, as shown in the Fig.(\ref{pairingvertex}):

 \begin{figure}[tbh]
  \centering
\includegraphics[width=0.6\textwidth]{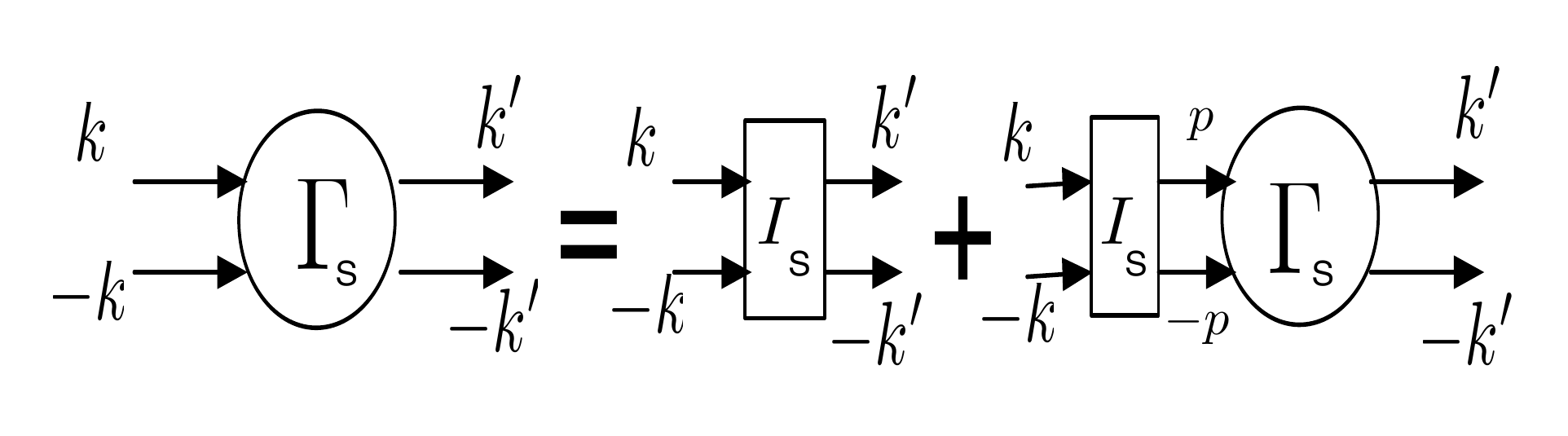}
\caption{Bethe-Salpeter Equation for the relation between the complete particle-particle vertex $\Gamma$ and the irreducible vertex $I$.}
\label{pairingvertex}
\end{figure}
\be
\label{BSEqn.}
\Gamma_{S}({k, k+q}) = I_{S}({k, k+q}) - (T/2) \sum_{{\bf p}, \omega} I_{S}({ k,p}) G({\bf p}, \omega)G({\bf -p}, -\omega) \Gamma_{S}({p, k+q}).
\ee
Here $I_S$ ({ k,p}) is the {\it irreducible} vertex in the particle-particle channel with total momentum zero, which means it contains all diagrams which cannot be cut into two parts by cutting two particle lines (right-ward going lines in the convention of Fig. (\ref{BSEqn.})). $G({\bf p}, \omega)$ is the single-particle Green function at momentum-energy ${\bf p}, \omega$. In (\ref{BSEqn.}), and fig.(\ref{pairingvertex}), the integral over energy has a cut-off energy of the vertex $I_S$. This cut-off $\omega_c$ is the upper energy scale of the collective fluctuations of the fermions or of other modes with which the fermions interact. For problems of interest, we expect this cut-off energy to be much smaller than $E_f$ but much larger than $T_c$.
This can in general only be justified {\it aposteriori}. 

Quite generally, for the e-ph or the e-e problem, one can define  a four-fermion interaction vertex $ I_{S}({k, k+q})$ which is irreducible in the particle-particle channel through the interaction Hamiltonian
\be
H_{int} = 1/2 \sum_{{\bf q}, \nu, \omega_n} \sum_{\bf k} g_{\alpha,\beta}^{\nu}({\bf k, k+q})g_{\gamma,\delta}^{\nu}({\bf -k, -k-q}) \mathcal{F}_{\alpha,\beta,\gamma,\delta}({\bf q}, \nu \omega_n)  c^+_{{\bf k+q}, \beta} c^+_{-{\bf k+q},\gamma} c_{-{\bf k}, \delta}c_{{\bf k},\alpha} 
\ee
$\mathcal{F}_{\alpha,\beta,\gamma,\delta}({\bf q}, \omega_n)$ is the propagator of the fluctuations which are exchanged by the fermions, $\omega_n$ sums over the boson Matsubara frequencies, and $g_{\alpha,\beta}({\bf k, k+q},\nu)$ is the interaction vertex with $\nu$ denoting any additional idex, such as the polarization for phonons, needed to specify the coupling to the fluctuations. For interactions with phonons or any other density or current fluctuations $\alpha = \beta; \gamma = \delta$. For isotropic interactions with spin-fluctuations or spin-current fluctuations, the spin-dependence in the fermion operators and the Fluctuation propagator in  has the form $\sigma_{\gamma,\delta}\cdot \sigma_{\beta,\alpha}$.

In general three integral equations specify the Eliashberg solution: the equation for the normal self-energy of fermions, the equation for the pairing self-energy of the fermions, and the equation for the self-energy of the Bosons due to superconductivity. The latter is unimportant for the e-ph problem because the corrections due to superconductivity are of order $(m/M)^{1/2}$. But they are likely to be very important for e-e problem below $T_c$, because superconductivity itself changes the spectrum of collective e-e fluctuations by gapping the low energy single-particle spectrum. This will have quite significant effects, for example on the temperature dependence of the 
superconducting gap. However, since we are interested here only in $T_c$ and that too only in a mean-field approximation, this will be neglected.

\subsection{Considerations on Pairing Symmetry}
\subsubsection{\bf Pairing Symmetry due to e-ph Interactions}

Let us start with the familiar example of electron-phonon interactions when the irreducible interactions are specified by 
\be
 H_{e-ph} = \sum_{{\bf k, q}, \nu}\sqrt{\hbar/(2M\omega_{{\bf q} \nu})}  I_{\bf k, q}^{\nu} c_{{\bf k+q}\sigma}^+c_{{\bf k}\sigma}(b_{{\bf q} \nu} + b_{-{\bf q} \nu}^+),
 \ee
 $b_{{\bf q} \nu}, b_{{\bf q} \nu}^+$ are the phonon creation and annihilation operators and  $g^{\nu}({\bf k, k+q}) = \sqrt{\hbar/(2M\omega_{{\bf q} \nu})}  I_{\bf k, q}^{\nu}$ is the electron-phonon vertex, $M$ is the ion-mass and $\omega_{{\bf q} \nu}$ are the frequencies of phonons of momentum ${\bf q}$ and polarization $\nu$. A specific form for the electron-phonon vertex will be introduced later.
 
Optical phonons are to a first approximation dispersion free. Their interactions with electrons in a metal are effectively $\delta$-function in real space, which only gives $s$-wave scattering. The interaction with acoustic phonons must vanish in the long wavelength limit. Moreover, the partial density of states of phonons is peaked near the zone-boundary. The effective interaction is therefore short-range, of the order of the lattice constant. As will be discussed later, this is true for pseudo-potential metals in which the e-ph interactions are weak. For transition metals and compounds and other co-valent metals with larger interactions and larger $T_c$, the e-ph interactions are even shorter-range. Dominant attraction is therefore always in the $s-$ wave channel. 

\subsubsection{\bf Effective electron-electron repulsion in the s-wave channel in Jellium}

Electrons also repel. 
The effective dimensionless electron-electron interaction (normalized to the density of states at the chemical potential), and assumed to be cut-off in energy only at the Fermi-energy, 
\be
\label{mu}
\mu = N(0) <4\pi e^2/(q^2 \epsilon(q))>
\ee
where $\epsilon(q)$ is the dielectric function and the average is over initial and final states at the chemical potential in the same manner as in the definition of the $\lambda$ in Eq. (\ref{I2}). It is interesting to consider the relationship of $\mu$ to $\lambda$ with both calculated for jellium.
The dimensionless electron-phonon coupling constant $\lambda$ for jellium, defined in analogy with (\ref{lambda}) is
\be
\lambda^{e-ph} = N(0)<g_{scr}^2(q)>/M <\omega(q)^2>,
\ee 
where $g_{scr}(q)$ is the screened electron-phonon coupling function, related to the bare el-ph interaction function $g(q)$ by
\be
g_{scr}(q) = g(q)/\epsilon(q),
\ee
and the average is over the same states as in \ref{mu}. For jellium, $g(q)$ is given by
\be 
g^2(q)/(M\Omega_p^2) = 4 \pi e^2 /q^2,
\ee
where the bare ion plasma frequency $\Omega_p$ is related to the actual frequency of the ions in jellium by the Bohm-Staver relation,
\be
\omega^2(q) = \Omega_p^2/\epsilon(q).
\ee
We therefore have \cite{cmv-conf}
\be
\lambda^{e-ph} = \mu.
\ee

In view of the fact that jellium is completely characterized by one parameter $r_s$, such a relation should not be surprising. One could go further and say that, this relation could be projected onto any angular momentum pairing channel $\ell$ of electrons scattering from ${\bf k, -k}$ to ${\bf k',-k'}$ and therefore that such a relation exists in each channel, $\lambda^{e-ph}_{\ell} \approx \mu_{\ell}$. For Fermi-Thomas screening $\epsilon(q) \propto q^2/(q^2 + q^2_{TF})$, where $q_{TF}^{-1}$ is the inverse screening length. One may project the interactions on to pairing channel $\ell$ and find that the effective dimensionless direct repulsion as well as e-ph induced attraction parameters both are proportional to $\exp(-{\frac{q_{TF}}{k_F}\ell})$. 

Cohen and Anderson \cite{Cohen-Anderson}, in a not easily accessible paper, which has several interesting ideas, have discussed how in a lattice, Umklapp scattering which in effect is due to the modulation of the charge density within a unit-cell due to phonons or local field effects, increases the coupling constant $\lambda^{e-ph}$ relative to  $\mu$. This paper emphasizes that strong e-ph interactions mean the modulation of electronic bonds. This point will be discussed further in Sec. (III) in the context of transition metals and compounds, where this idea is most prominently borne out in the empirical data and implemented efficiently through the tight-binding representation of e-ph interactions to provide a theory of e-ph interactions as well as a quantitative theory of the anomalies in the phonon spectra \cite{cmv-weber, cmv-eph} 
in transition metal and compounds.

We should also note the important fact that when the effective e-e interactions in the pairing channels are expressed so that they are retarded only over the same range, $<\omega>$, as the e-ph interactions they are reduced to 
\be
\mu^* = \mu/(1+ \mu \ln(E_f/<\omega>).
\ee

\subsubsection{\bf Pairing in $\ell \ne 0$ through Incoherent particle-hole fluctuation exchange}

In an interesting paper Kohn and Luttinger \cite{kohn-lutt} showed that the effective electron-electron interactions must, due to the sharpness of the fermi-surface, always be oscillatory in the momentum transfer $2k_F\cos(\theta)$ where $\cos (\theta) = {\bf \hat{k}_f}\cdot {\bf \hat{k}'_f}$, with both ${\bf k}$ and  ${\bf k}'$ are close to the fermi-surface. From this it follows that there is always attraction in the pairing channel at some or the other angular momentum angular momentum $\ell$. The issue then is whether it is of sufficient magnitude to give $T_c$ larger than that from typical e-ph interactions.

Kohn and Luttinger also provide an estimate for the effective interaction when the bare interaction is weak. The direct (screened for e-e interactions) repulsive interaction falls of  exponentially in $\ell$, while the oscillatory exchange  attractive interactions fall off only as $1/\ell^4$, the latter therefore wins for large enough $\ell$. The same is true for hard-core interaction of finite radius. For $He^3$, taking this radius equal to the diameter of the He atoms, $\ell =1$ channel is already attractive and provides using a BCS type expression, a $T_c \approx \omega_{c} \exp (-2.5)$. Here  $\omega_{c}$ is the cut-off. If one takes $\omega_c \approx E_f$, one gets a factor $10^2$ too large a value compared with the experiments. There are several things wrong with this. For $\ell \ne 0$ pairing, one must include the repulsion in the $\ell=0$ channel in the normal self-energy, even in the weak-coupling limit. This cuts down the estimate by about two orders of magnitude. The leading effect over the weak-coupling limit is that, the quasi-particle renormalization, due to increasing interactions, also renormalizes the attractive coupling constants in the exponent downwards as well as the cut-off $\omega_c$. 

An important point is that for any reasonable interactions, the Kohn-Luttinger result that the attractive interaction falls off as $1/\ell^4$ is unlikely to change much in more complicated calculations with incoherent particle-hole interactions. Then, for example, for the parameters of $He^3$, the $\ell=2$ pairing has a transition temperature of about $10^{-17} K$, even without considering the s-wave repulsion.
One may conclude that incoherent particle-hole fluctuations are not a very good way to get to high $T_c$. How then to make use of the higher cut-off energy of electronic excitations. The alternative to incoherent particle-hole fluctuations are collective modes of electronic fluctuations, to which we turn next. 

\subsubsection{\bf Exchange of collective fluctuations}

Part of the spectral weight of particle-hole fluctuations appears in collective modes if the effective interactions are strong enough compared to the kinetic energy. In general a particle hole-fluctuation is characterized by an internal momentum ${\bf k}$ and a center of mass momentum ${\bf q}$, besides other indices which specify 
the channel of the excitations, for example, density or spin or current or spin-current or inter-valley, interband, etc. A useful limit is obtained from the fact that a bound state of a fermion particle and hole (and indeed particle and particle or hole and hole) has Boson commutation relations. In a bound state, there is no dispersion in energy at any ${\bf q}$ as a function of ${\bf k}$, so the latter index may be summed over. To regard the fluctuations as collective rather than incoherent, one must make sure that  the dependence on ${\bf k}$  is unimportant  to do sensible calculations. 
Also important is the integrated spectra of the collective modes and the incoherent particle-hole spectral weight of fermions for any given channel, is generally fixed by sum-rules. It is sinful double-counting to 
introduce collective variables, for example spin-fluctuations, and give them the total weight of $S(S+1)$ per spin $S$, and do calculations of their interactions with fermions in the same way that one does for phonons.

\subsubsection{\bf Symmetry of Pairing Interaction from Exchange of Spin-Fluctuations}

 As we will see later, the frequency dependence of the interaction (with total spectral weight and coupling constant fixed) is of much more importance in determining $T_c$ in finite $\ell$ than for $\ell =0$. The anisotropy of the interaction as well as the details of the geometry of the fermi-surface(s) are also important in determining the irreducible representation favored for pairing. However, essential aspects of  finite $\ell$ pairing, specifically the distinction between the pairing promoted by ferromagnetic and anti-ferromagnetic fluctuations is revealed in a simple model calculation \cite{msv} based on a frequency independent and isotropic interaction in a model of a spherical fermi-surface. Consider effective electron-electron interactions of the simple form:
\be
\label{modelSF}
H_{int} = 1/2 \sum_{{\bf kk'q}}  J(|{\bf k-k'}|) \sigma_{\alpha \beta}\cdot \sigma_{\gamma \delta} c^+_{{\bf k + q/2}, \alpha} c^+_{{\bf -k + q/2}, \gamma} c_{{\bf -k' + q/2}, \delta} c_{{\bf k' + q/2}, \beta}
\ee
In the weak-coupling approximation, the partial wave components of the pairing interaction $V_{\ell}$ follow from Eq.(\ref{modelSF}) to be
\be
\label{Vell}
V_{\ell} =  \left(
  \begin{array}{c}
    3\\ 
    -1 \\ 
  \end{array}\right) 2 \int_0^1 dx x P_{\ell}(1-2x^2)[- J(2k_F x)],
\ee
where the pre-factor is $3$ for spin-singlet (even $\ell$) pairing and $-1$ for triplet-spin (odd $\ell$) pairing. 

\begin{figure}[ht!]
  \centerline{
  \includegraphics[width=0.5\columnwidth]{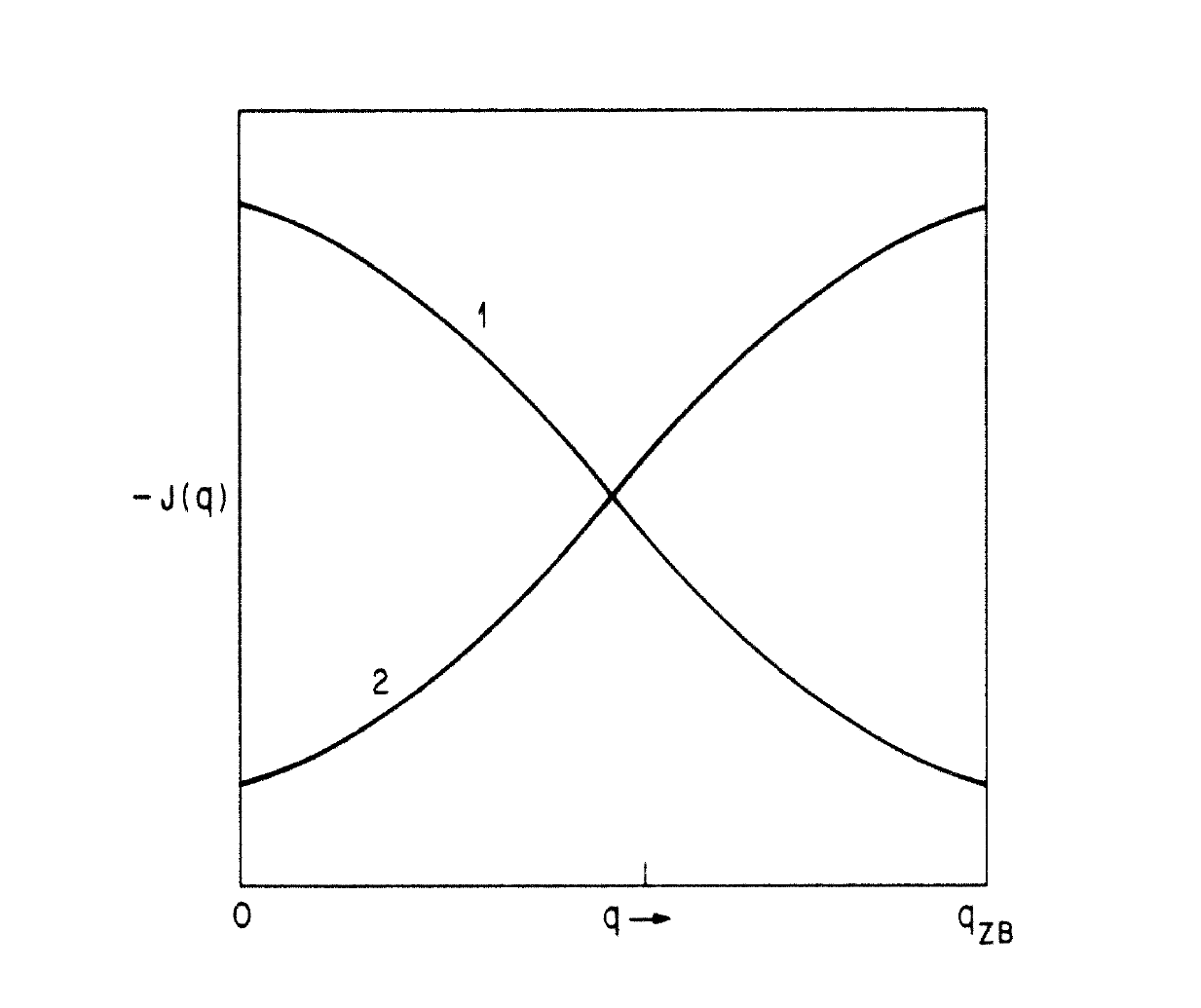}}
  \caption{q-dependece of FM and AFM interactions} \label{FMAFM}
\end{figure}

One may analyze the consequences of the $q$- dependence of $J(q)$ in of Eq.(\ref{Vell}) using fig. (\ref{FMAFM} ). The ferro-magnetic interactions have a peak of $-J(q)$ at $q = 0$, while the anti-ferromagnetic interactions have a peak at the zone-boundary. First note that if $J(2k_F x)$ is independent of $x$, $V_{\ell} =0$ for all $\ell$. This is because $J(q)$ independent of $q$ implies a delta-function interaction in real space where any finite $\ell$ Cooper pair wave-function has zero magnitude. On the other hand for $\ell = 0$, a constant $J(q)$ represents a magnetic impurity, which suppresses pairing. One also can see from Eq.(\ref{Vell}) and the smooth variation of 
$J(q)$ in fig. (\ref{FMAFM}) that $V_0 > 0,  V_1 < 0$ and $V_2> 0$ for ferro-magnetic interactions. In other words both isotropic and anisotropic singlet states are disfavored by such interactions, while triplet $\ell=1$ state is favored. For anti-ferromagnetic interactions, $V_0 > 0, V_1 >0$; both isotropic singlet and the triplet state is disfavored. Whether 
$V_2 <0$ or $ >0$, depends on the detailed form of $J(q)$. A strong enough peak in $J(q)$ near the zone-boundary favors such pairing.

It is straightforward to extend this analysis \cite{msv} to the more realistic case for the momentum dependence of the interaction taking the crystal symmetry into account. It is expected that the antiferromagnetic fluctuations peak at a point ${\bf q = Q}$ along a symmetry axis and so does $J({\bf q})$. I refer to the original paper for some details. An important point to emerge from that analysis is that the pairing symmetry depends both on how steep is the increase in $J({\bf q})$ near the zone-boundary and on the details of the electronic dispersion $\epsilon({\bf k})$ near the chemical potential. For example the electronic structure may choose between {\it extended - s} or d-type pairing. This point appears relevant to the case of the recently discovered Fe-pnictide superconductors, where the band-structure with hole-pockets at the zone center and electron pockets near zone corners may favor nodeless pairing with a phase difference of $\pi$ between the gaps at zone-centers and the gaps at zone corners. It appears possible that for appropriate band-structure AFM fluctuations may even promote triplet pairing. This may be implicated in the triplet superconductivity in $Sr_2RuO_4$ \cite{maeno-sr2ru04}, which shows no FM correlations but modest AFM correlations.

One can physically understand for the simple square lattice band-structure why a steep increase of $J({\bf q})$ near the zone-boundary favors d-wave. The wave-function for this interaction favors anti-parallel correlations of spins on nearest neighbors. Such a spin-correlation is also produced by the $d-$wave BCS wave-function, for such a band-sturcture.

A generalization of magnetic fluctuations due to spin-moment correlations to magnetic fluctuations due to orbital-moment correlations will be discussed in connection with the physics of the cuprates in Sec. (IV).

While the arguments above may provide valid grounds for discussing pairing symmetry, they are no help in thinking about the value of $T_c$. For that one must turn, as we do next, to the frequency dependence of the pairing vertex, which in most cases of interest is given by the frequency dependence of the collective modes exchanged by the fermions.

\subsection{\bf Effect of the Frequency Dependence of Fluctuations on $T_c$ in $s$-wave and higher angular momentum pairing}

We know that we do need for high $T_c$:\\
\noindent
 (A) Large spectral weight in the collective fluctuations being exchanged by the fermions, \\
 (B) Large coupling of these fluctuations to the fermions, \\
 (C) Large frequency scale of the fluctuations and \\
 (D) Large density of states of fermions around the chemical potential in a range of frequencies of the upper cut-off of the fluctuations. \\

 (A) is fixed for phonons but not for e-e fluctuations and must be carefully considered so as not to overcount the degrees of freedom and for its effect in renormalizing (B). We will see in Sec. III that (B), (C) and (D) are not independent in the e-ph problem. Arguments will be given that this is true also for the e-e problem.  For the case of e-e interactions there is also the additional and important consideration of the distribution of the spectrum of fluctuations, due to the role of inelastic scattering and the difference of normal and anomalous self-energy. We proceed to discuss this immediately. 
  
 To discuss the role of the spectrum of fluctuations one must turn to the careful analysis of the solutions to the Eliashberg equations. This was done in a beautiful paper for s-wave pairing through phonons by Bergmann and Rainer \cite{bergmann-rainer}, followed up to examine the role of AFM fluctuations in s-wave pairing \cite{tvr-cmv} and finally for d-wave pairing \cite{mscmv}.
 
 Let us start with the simplest situation when the spectrum of fluctuations is extended more or less uniformly over a large range up to a cut-off $\omega_c >> T_c$.  A simple generalization \cite{ mscmv, levin-valls} of the McMillan approximate solution \cite{mcmillan, allen-dynes} to $\ell \ne 0$ pairing gives
 \be
 \label{tc}
T_c \approx \omega_c \exp{(\frac{-1 + \lambda_s}{\lambda_{\ell}})}
\ee
for $\lambda_s, \lambda_{\ell}$ of $O(1)$ or smaller. $\lambda_{\ell}$ is the dimensionless coupling constant in the $\ell$-th particle-particle channel in which pairing is presumed to occur and which therefore occurs in the pairing self-energy in the Eliashberg equation. 
Since the self-energy must respect the full symmetry of the lattice this always includes the $\ell =0$ channel. For a square lattice in two dimensions or a cubic lattice in three dimensions, the first $\ell \ne 0$ momentum channel included in $\lambda_s$ is the fourth. Since in any reasonable model, the coupling constants go down fast with $\ell$, we need only to include the $\ell =0$ in it. Eq.(\ref{tc}) has the consequence that for short-range interactions, $T_c/\omega_c$ is smaller for $\ell \ne 0$ than for $\ell=0$, with the decrease depending on the range of interactions. 

Just as Eq.(\ref{tc}) does not include the effect of pair-breaking due to any magnetic impurities present for $\ell =0$ wave or due to any form of impurity for $\ell \ne 0$, it does not include the effect of pair-breaking due to inelastic (real processes as opposed to virtual processes) scattering. These processes weigh the spectrum of fluctuations over the range that their frequency is smaller than $O(T)$. More detailed analysis of the Eliashberg equations is required to study this.

 A quantity of interest to study the role of the frequency dependence of the fluctuations is the functional derivative of $T_c$ with respect to the change in the spectral function $A(\omega)$ \cite{bergmann-rainer} which appears in the kernel of the Eliashberg integral equations: 
\be
\label{dTc}
\frac{1}{\omega} \frac{\delta T_c}{\delta A(\omega)}
\ee
For s-wave pairing, the analysis of Bergmann and Rainer shows that $ \frac{\delta T_c}{\omega\delta A(\omega)} > 0$ for any frequency for s-wave pairing, reducing to $0$ in the limit $\omega \to 0$, consistent with Anderson's theorem about the effect of weak impurity scattering on $T_c$. Increase of spectral weight in any region of the frequency spectrum increases $T_c$ for s-wave pairing. The maximum value of this quantity appears to occur near $\omega \approx 2 \pi T_c$.

\begin{figure}[ht!]
  \centerline{
  \includegraphics[width=0.5\columnwidth]{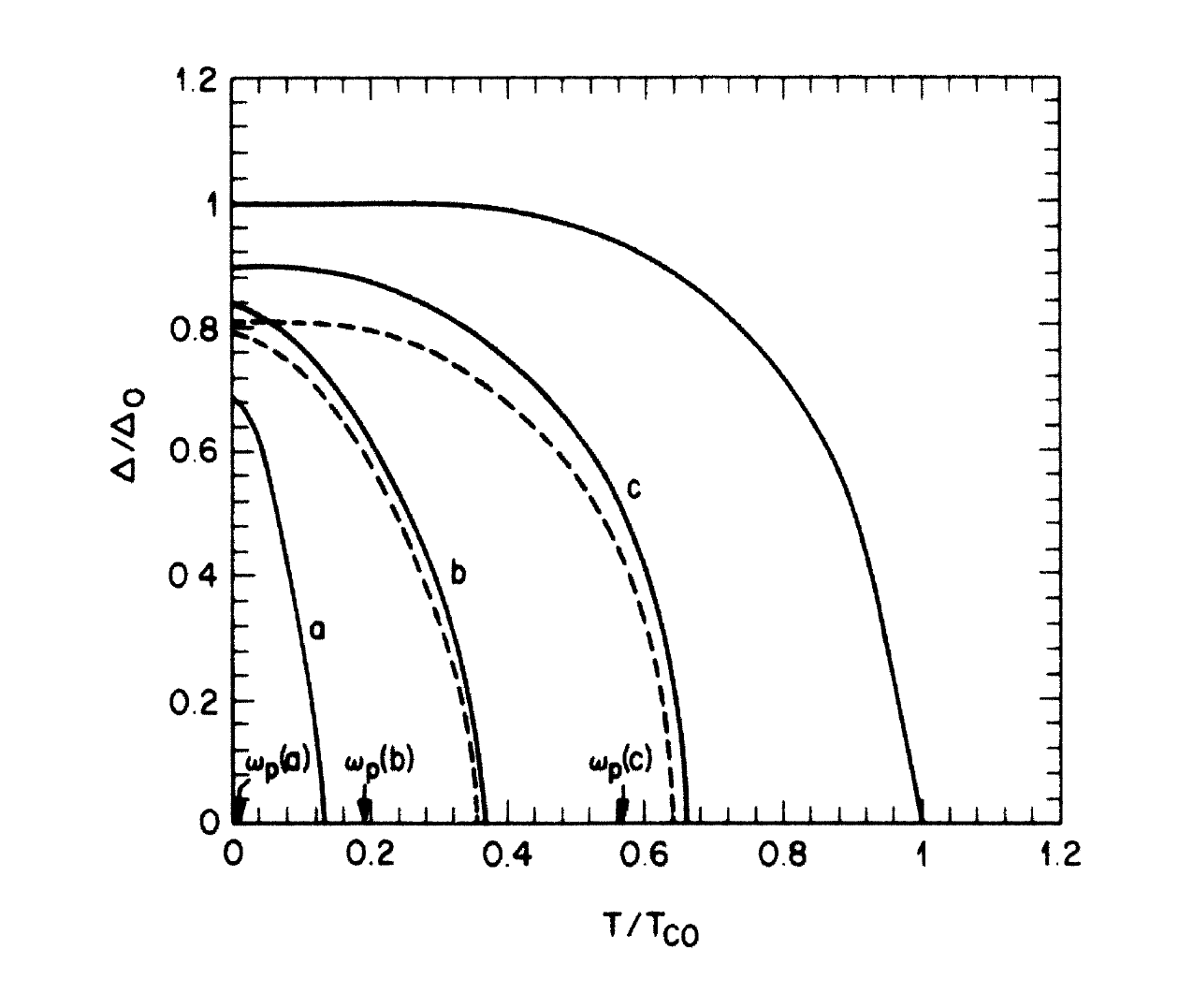}}
  \caption{Dependence of $T_c/T_{c0}$ and $\Delta/\Delta_0$ on inelastic scattering for the case of d-wave pairing. An Einstein spectrum of spin-fluctutations at $\omega_E$ is used for simplicity with frequencies 1 and additional Einstein modes are introduced with frequencies (in units of $\omega_E$) as marked for curves a, b and c. $T_{c0}$, and $\Delta_0$ are the values without inelastic scattering and so is the unmarked curve. Dashed and full lines represent approximate and exact solutions of the Eliashberg equations. For details, see (\onlinecite{mscmv})}
  \label{Tc/Tc0}
\end{figure}

For finite $\ell$, the situation is different and depends on the ratio $g = \lambda_{\ell}/\lambda_s$. Then, for $g < 1$, one can define a frequency \cite{mscmv}
\be
\omega^* \approx T_c \exp (1/g)
\ee
such that for $\omega \gtrsim \omega^*,~ \omega \frac{\delta T_c}{\delta A(\omega)} \gtrsim  0$, while for $\omega \lesssim \omega^*,~  \omega \frac{\delta T_c}{\delta A(\omega)} \lesssim  0$. 
The physical reason for this is that thermal occupation of the fluctuations (other than those of long wavelenth which only lead to forward scattering) has similar effect as impurity scattering. This is pair-breaking for $\ell \ne 0$ because it mixes the phase of gap at different parts of ${\bf k}$-space in the pure limit.  
These conclusion are found in the solution of the Eliashberg equations for an Einstein model of spin-fluctuations, see Fig.(\ref{Tc/Tc0}). Note, also the much larger reduction in $T_c$ compared to the zero-temperature superconducting gap $\Delta(0)$, leading to a large value of $\Delta(0)/T_c$ characteristic of the Cuprates. 

\subsection{\bf Quantum-Criticality in Relation to $T_c$ in e-e Induced Pairing}

 The highest $T_c$'s in the Cuprates are undoubtedly through electronic fluctuations, and near a quantum-critical point (qcp), as are those in the heavy-fermions. This as we will see is likely be true for the pncitides as well. 
As discussed below for each of these cases, we know this because the observed transport and thermodynamic properties for $T \geqslant T_c$ in these compounds can only be understood as due to scattering of fermions from fluctuations which have a singularity in the limit $\omega \to 0, T \to 0$. It is therefore important to ask about the role of spectral distribution of fluctuations near qcp's in determining $T_c$ in light of what we have learnt in the previous section. 

Much less is understood about the universality classes of quantum critical points (qcp) than classical critical points where the scale-invariant frequency and momentum dependence of critical  fluctuations for the classical critical points has been catalogued into different Universality classes \cite{hohenberg-halperin}.
A qcp is the point $p = p_c$, where by varying a parameter $p$, for example pressure or electron density, the transition temperature to some broken symmetry $\to 0$.  A simple generalization of the dynamics of classical critical phenomena to the quantum case \cite{bealmonod-maki, hertz, moriya, rosch-rmp, cmv-phys-rep-sfl} (which may be termed a Gaussian qcp since the fluctuation spectra is determined by the renormalized coefficient of the quadratic term of the variable describing the fluctuating order parameter) is as follows: The spectral function of the fluctuations near a qcp may in general be be written in the scaling form:
\be 
\label{qcp-flspectra1}
\chi "(q, \omega, p, T) = \chi_0 \xi^{-d_M} \mathcal{F}_1\big(k\xi, \omega \xi^z, \omega/T\big),
\ee
$\xi_r$ is the correlation length at $T \to 0$ which diverges as
\be
\xi_r \propto |p-p_c|^{-\nu},
\ee
and $k$ is measured from the wave-vector of the symmetry breaking.
$d_M = d$, the physical dimension if the order parameter is that of a conserved quantity, otherwise it can be different.
The correlation length in the time-direction or equivalently the frequency scale of the fluctuations  is given by
\be
\xi_t \propto \xi_r^z.
\ee
$z$ is the dynamical critical exponent which scales the frequency of fluctuations to their characteristic spatial extent reflecting that the dispersion of the critical fluctuations depends on the nature of the broken symmetry and in general has the form $\omega \propto k^z$.
Eq.(\ref{qcp-flspectra1}) differs from that in classical dynamical critical phenomena in only two ways: the substitution of $(p-p_c)$ at $T=0$ for $(T-T_c)$ and that the fluctuation frequencies $\omega$ also has a scale which is simply the temperature of measurement $T$, i.e. the distance along the frequency axis from the quantum-transition at $(p=p_c)$ and $T=0$. These two ensure that there
 is a scale $T_x(p)$  which gives the crossover from quantum fluctuations at low T to classical fluctuations at high T and which marks the temperature below which fermi-liquid properties are expected. As opposed to classical critical phenomena, the critically-scaled behavior in physical properties is to be expected in the entire regime bounded by the transition temperature $T_c(p)$ and the crossover scale $T_x(p)$ and the upper cut-off temperature of the fluctuations, given by the microscopic energy scales (for example, the exchange splitting in a ferro- or antiferro- qcp). This is usually referred to as the quantum-critical regime of the phase diagram.
 
It is useful to take $\xi_r, \xi_t$ as dimensionless, the former normalized to a lattice constant and the latter normalized to the upper cut-off of the fluctuations $\omega_c$.  Eq. (\ref{qcp-flspectra1}) has been written so that $\chi_0$ is the spectral weight of the critical fluctuations near the critical point, by which I mean that $\chi"(q,\omega, p, T)$ integrated over all $\omega$ and $k$ is $\chi_0$. The system of-course has other fluctuations but $\chi_0$ saturates only part of the total spectral weight.
 
 The form (\ref{qcp-flspectra1}) is suitable for discussing experiments as a function of $(p-p_c)$. For analyzing experiments as a function of frequency and temperature, an alternate form may be more useful:
 \be 
\label{qcp-flspectra2}
\chi "(q, \omega, p, T) = \chi_0 T^{-d_M/z} \mathcal{F}_2\big(\frac{k}{T^{1/z}}, \frac{\omega}{T}, \frac{1}{T\xi_t}\big).
\ee
 Thermodynamic and transport properties near a Gaussian qcp depend on the cut-off scale $\omega_c$, the spectral function $\chi_0$, the temperature $T$ and the exponent $z$. The conventional theory of quantum criticality of ferromagnets gives $z=3$ for ferromagnets and $z=2$ for antiferromagnets. In this situation,
 the frequency scale as well as the momentum scale of the fluctuations $\to 0$ as the critical point is approached just as near classical critical points; at a temperature $T$ at $p=p_c$, the frequency-scale of the fluctuation is $T$, and given in terms of $\xi_t$ by (\ref{qcp-flspectra2}) away from it. 
   
 Let us look at what such critical fluctuations do for $T_c$ through what we have just learnt in the previous section. In the quantum-critical regime, the scale of fluctuations just above $T_c$ or the cut-off scale in the fluctuations exchanged is $T_c$ itself. This is bad for two reasons: the prefactor of the expression for $T_c$ goes down with the cut-off, and as discussed and illustrated for $\ell \ne 0$ pairing in Fig.(\ref{Tc/Tc0}), inelastic scattering has a particularly bad effect on $T_c$. 
 
Is this compensated for by increase in the dimensionless coupling constants $\lambda_{\ell}$? In other words does the peaking of fluctuation at low frequencies and in a region of width $\xi_r^{-1}$ around the critical wave-vector increase the coupling constants. The answer is sadly, no. The point is that the coupling constant depends on the integrated value of the fluctuations over $q$ and $\omega$ and these are essentially fixed by sum-rules. There is a weak (logarithmic) increase  in the coupling constant if the spectra peaks at low frequencies which is generally quite uninteresting. 

Let us consider the role of $\chi_0$, the spectral weight. It may be taken from the ordered moment $<M>$ far away from $p_c$, where it is a slow function of $p$, to be $\approx <M>^2$. This is a fraction $\bar{\chi}_0$ of the totals spectral weight of such fluctuations, the rest remaining as incoherent particle-hole fluctuations of fermions. As discuss below, the irreducible vertex is reduced by $\bar{\chi}_0$ compared with the case for phonons.
 
 In Sec. IV, we will review the firm experimental evidence for the case of the Cuprates and for the Heavy-Fermions that the qcp is not of the Gaussian kind. The fluctuations appear to be local in space and decay as power laws in time. In other words $\chi "(q, \omega, p, T)$ at $p=p_c$ has no singularity as a function of $q$ but has a singularity as $\omega$ and $T \to 0$. The upper cut-off in the fluctuations is at and energy $\omega_c$, which is only about an order of magnitude smaller than $T_c$. Critical fluctuations  proposed on phenomenological grounds for the Cuprates have the form $\chi "(q, \omega, p_c, T) = \chi_0 \tanh (\omega/T)$, with a cut-off at $\omega_c$. One could formally put a dynamical exponent $ z \to \infty$ and put this in the form in Eq.(\ref{qcp-flspectra2}). But this obscures that quite different physical ideas are required to derive such fluctuations than those of the Gaussian kind. For example, the derivation of this class of fluctuations (\cite{aji-cmv-qcf}, \cite{aji-cmv-qcf-pr}) for Cuprates rests on finding  two classes of orthogonal topological variables one of which has spatially local and temporally logarithmic interactions while the other has spatially logarithmic interactions and temporally local interactions.  Classical statistical models are known where the singular fluctuations are determined by topological excitations, for example the Kosterlitz-Thouless transition in the 2d x-y model and transitions in several vertex models \cite{baxter}, which do not fall under the purview of the classical model of phase transitions which are the inspiration for the theory of Gaussian qcp's. The class of microscopic models where such quantum-criticality occurs is not known. I will refer to such qcp's as topological qcp's.
 
 The absence of a spatial scale and the freeing of the scale of frequency of the critical fluctuations from the requirement that they tend to lower values as $T \to 0$, removes the two deleterious effect of the Gaussian critical fluctuations on $T_c$: the prefactor in the expression for $T_c$ remains at $\omega_c$ and there is essentially no extra pair-breaking due to inelastic scattering for $\omega_c >> T_c$. The locality of the spatial scale, which in practice means that the spatial scale is similar to $k_F^{-1}$ has the additional consequence that the $\ell=0$ and $\ell \ne 0$ couplings are similar in magnitude. In fact, given a total spectral weight $\chi_0$ and the $\ell$ in which pairing is favored, it is not possible within Eliashberg theory to have a better spectrum for high $T_c$ than such a spectrum.
  
 


{\bf D. Calculations of $T_c$ for the Hubbard Model}

Following the suggestion by Anderson \cite{pwa-science} that the essential physics of the Cuprates is  described by the Hubbard model, there have been innumerable attempts by a variety of methods to calculate the properties of the Hubbard model, including $T_c$. The approximate methods (RPA \cite{monthoux}, varieties of dynamical mean-field theories \cite{jarrell, tremblay}, variational Monte-carlo \cite{ogata-vqmc, sorella-vqmc}) in comparison with the best numerically precise method, Monte-Carlo without sign problems for $U/t$ up to 7, \cite{Imada-qmc}, illustrate the hazards of the enterprise. While the last give the upper limit, $T_c /t < 10^{-3}$, the various approximations give a value one and sometimes two orders of magnitude larger. 

Since (apart from the RPA), such calculations present results for $T_c$ without providing the form of the fluctuations spectra or of the coupling function or calculate normal state properties, it is hard to say what exactly goes wrong in even such elaborate and careful calculations. A hint for one of the possible deficiency from the best Monte-Carlo \cite{Imada-qmc} calculations done at various sizes of the lattice (up to $10 \times 10$) is that the nearest neighbor static spin-correlations of the d-wave superconducting wave-functions on a square lattice match those of the AFM wavefunction. So, it is possible for calculations on clusters of small size to lead to misleading conclusions which disappear for large enough size. 

In a later section, we will review the results of s-wave pairing in the attractive Hubbard model, where reliable results without numerical difficulties are available in special limits.

{\bf E. Excitonic and similar Pairing}

Noting the prefactor in the BCS expression for $T_c$, it is tempting to suggest attractive interaction among electrons near the fermi-surface through exchange of fluctuations whose frequency is high unlike the case of phonons where it is limited by $M^{-1/2}$, where $M$ is the ionic mass \cite{little, allender-bardeen, ginzburg}. (Note that there is no factor of $M$ in the coupling constant $\lambda$.) These other excitations must then be fluctuations of electronic degrees of freedom themselves or photons. For the latter, the coupling constant $\lambda$ is inevitably proportional to the fine-structure constant. So we need consider only the electronic excitations.  It seems that starting with Little \cite{little} no known excitation has not been thought of in this context. Although some of the ideas are not correct to begin with, most are in principle correct; the problem is in the smallness of the coupling constant. When they are wrong in principle, the mistake is of the following kind. Consider electron-electron interaction to second order in the Coulomb interaction: it is attractive. But if the whole series is summed, it gives just the screened repulsive Thomas-Fermi interaction. An example of this in connection with a proposal \cite {allender-bardeen} was provided by Inkson and Anderson \cite{inkson-pwa} and further checked through detailed calculations \cite{cohen-louie}. The original proposal by Little \cite{little} to use the excitons of an insulating side-chain in an organic metal has similar problems. 

It should be noted however that essentially every e-e mechanism we can think of is an excitation in the particle-hole channel and that if we get a decent $T_c$, it is generally always connected with a high frequency cut-off of the fluctuations. So all e-e processes we consider may be called Excitonically induced pairing. But all empirical information on high $T_c$ induced through e-e processes indicates that the 'excitons' must be particle-hole fluctuations of the same one-particle excitations that pair up and such that they engender singularities in the single-particle spectra so that the distinction between particle-hole fluctuations of the fermions and the 'excitons' is lost.

\section{Electron-Phonon interaction promoted $T_c$}

\subsection{ McMillan's Expression for the Coupling constant}

For a general spectral function,  
\be
B_{\alpha,\beta,\gamma,\delta}({\bf q},\nu, \Omega) \equiv \int_{-\infty}^{\infty} d\omega \frac{\mathcal{F}_{\alpha,\beta,\gamma,\delta}({\bf q}, \nu, \omega)}{\Omega-\omega},
\ee
the Eliashberg equations require a numerical solution. For the limited purposes of this paper, it is enough to use the McMillan simplified solution \cite{mcmillan}, where by the transition temperature in the s-wave channel is given for the e-ph problem by
\be
\label{Tc}
T_c \approx c \exp(-(1+\lambda^{e-ph})/\lambda^{e-ph}),
\ee
with 
\be
\label{lambda}
\lambda^{e-ph} \equiv \frac{N(0)<I^2>}{\kappa},
\ee
where $N(0)$ is the density of states of un-renormalized electrons at the fermi-surface, $<\omega>$ is an average over the phonon frequencies, $\kappa$ is a (suitably defined) average over the lattice stiffness, $\kappa = M<\omega^2>$, and $<I^2>$ is the scattering averaged over the spectral function of the phonons:
\be
\label{I2}
<I^2> = \frac{\int dS_{{\bf k}}\int dS_{{\bf k}'} \sum_{\nu} | g_{\bf k, q}^{\nu}|^2 v_{{\bf k}}^{-1}v_{{\bf k}'}^{-1}}{\int dS_{{\bf k}}\int dS_{{\bf k}'}v_{{\bf k}}^{-1}v_{{\bf k}'}^{-1}}
\ee
where a spin-trace (in the singlet channel) of $g_{\alpha,\beta}^{\nu}({\bf k, k+q})g_{\gamma,\delta}^{\nu}({\bf -k, -k-q}) $ has been taken to obtain  $| g_{\bf k, q}^{\nu}|^2$.

It is important to note that the $\lambda$ of Eq.(\ref{lambda}) also gives the mass renormalization in the normal state due to phonons so that the renormalization of the specific heat coefficient due to interaction with phonons is 
\be
(m^*/m)_{ph} = (1+\lambda^{e-ph}).
\ee
The coefficient of resistivity due to electron-phonon interactions in the normal state is also given in terms of $\lambda$. For example for temperatures comparable to an larger than $\omega$, the resistivity due to e-ph interactions in a metal may be written in terms of the scattering rate 
\be
\tau^{-1} (T) = 2\pi \lambda^{e-ph} T.
\ee

\subsection{Empirical Relations in transition metal superconductivity}

Study of empirical relations among the parameters determining Transitions temperatures in superconductors gives useful insights to the physics of metals.
\begin{figure}[ht!]
  \centerline{
  \includegraphics[width=0.7\columnwidth]{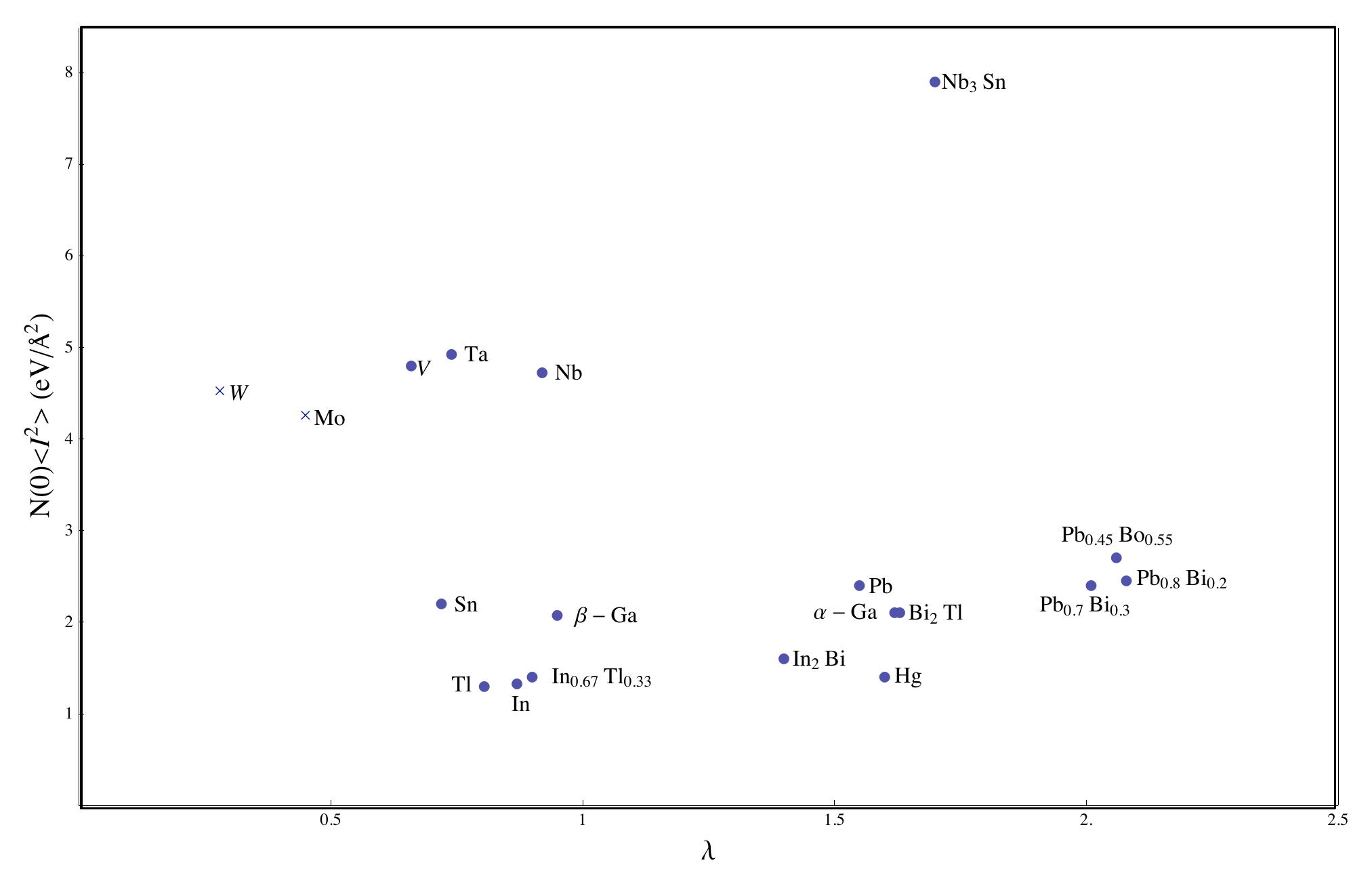}}
    \caption{Empirically determined "constancy" of $N(0)<I^2>$ in classes of metals and compounds.}
     \label{N0I2}
\end{figure}
In his analysis of the properties of superconducting metals and compounds, McMillan \cite{mcmillan} noted that $N(0) <I^2>$ varies only by about factor of $2$ while  $N(0)$ and $<I^2>$ vary by a factor of about $10$. However the empirically "constant" $N(0) <I^2>$ has different values for different classes of metals and compounds. For example, see Fig. (\ref{N0I2}), it is
 is close to one value for the pseudo-potential metals like Sn, Pb, Bi and their alloys and another for the transition metals and their alloys and yet another in the A-15 compounds. McMillan proffered no explanation for the transition metals but showed using the fact that the ion-plasma frequency $\Omega_p$ are always much larger than the phonon frequencies, that for the pseudopotential metals that $N(0)<I^2>/\Omega_p^2$ is approximately constant. 

Barisic, Labbe and Friedel \cite{friedel2}  presented a simple and strong argument for transition metals and compounds on the basis of the tight binding representation of the band-structure and of the electron-phonon coupling that $N(0) <I^2>$ is related simply to the cohesive energy $E_c$ of the metal. The argument is summarized in appendix A with the conclusion that 

\be 
N(0)<I^2> \approx N(0) <d^2E_c/dR^2> \approx N(0) E_c  /r_0^2.
\ee
$<d^2E_c/dR^2>$ is the average of the second derivative of the change in kinetic energy of the metal as the nearest neighbor distance between two atoms $R$ is changed leaving the others fixed; $r_0$ is the size of the typical metallic orbital.

\begin{figure}[ht!]
  \centerline{
  \includegraphics[width=0.7\columnwidth]{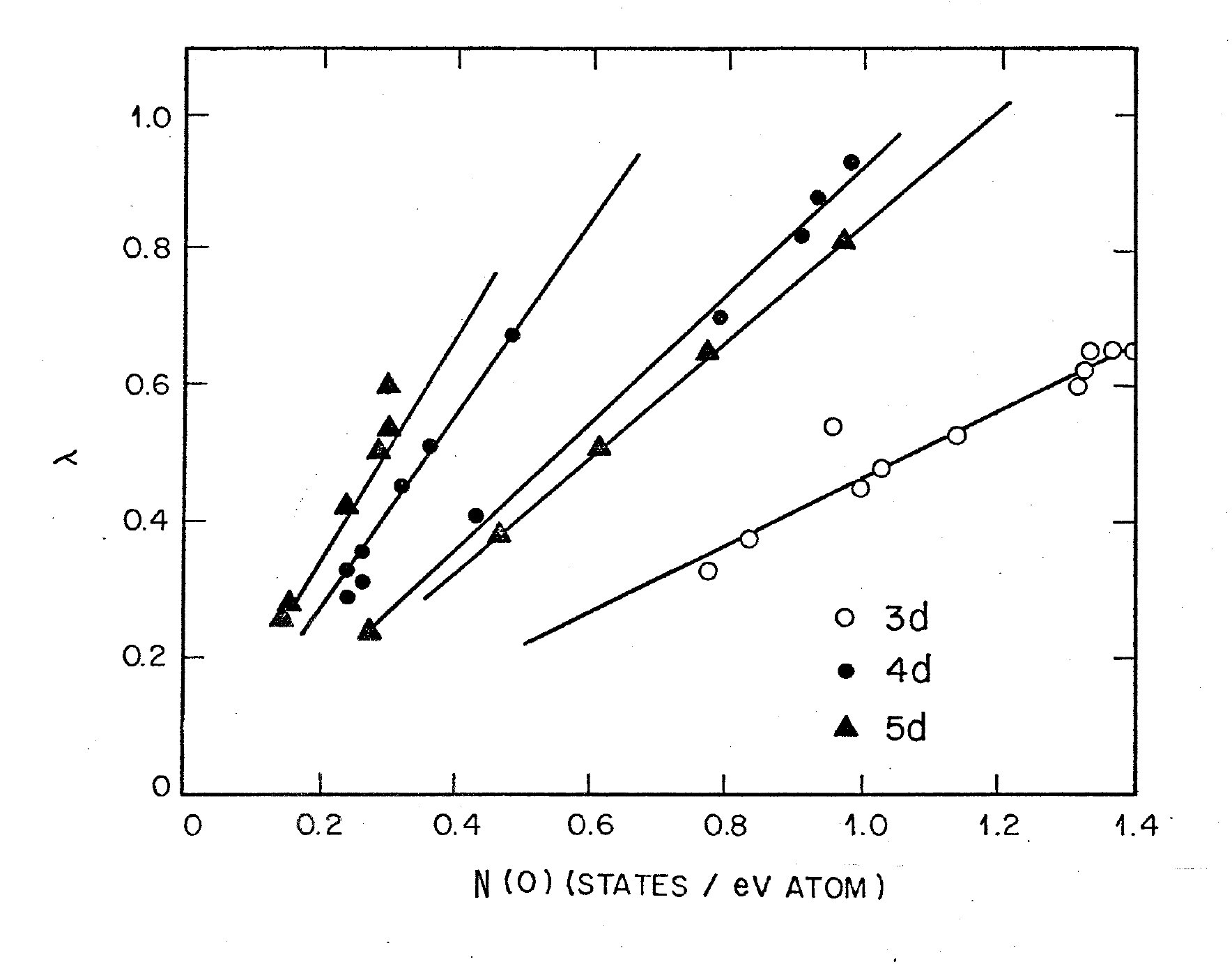}}
  \caption{Empirical Relation between experimentally deduced $\lambda$ and the bare electronic density of states at the chemical potential in 3d-4d and 5d metals and their alloys.} \label{I2/w2}
\end{figure}

One may be tempted to conclude that since $N(0)<I^2>$ within a given class of transition metals or compounds is approximately a constant, one may simply increase $\lambda$ by reducing the average lattice stiffness $M<\omega^2>$ and thus increase $T_c$. This led to the soft-phonon myth, much propagated in the 1970's . Quite apart from the fact that the prefactor $<\omega>$ would prefer matters the other way for high $T_c$, there is also another empirical rule \cite{hume-Blaugher, cmv-dynes} followed. It is that within a given class of materials $<I^2>/(M<\omega^2>)$ is also approximately a constant. 
For transition metals and alloys this is exhibited in fig. (\ref{I2/w2}). Only the first half of the 3d metals is superconducting; for 4d and 5d metals the two lines correspond to the fermi-level in the bonding part of the band and in the anti-bonding part of the band. This "constancy" with a different value is also followed in the A-15 family of superconductors \cite{cmv-dynes}. This rule may also be derived within the tight-binding approximation \cite{cmv-conf} and is summarized in the appendix.The final result is that for $N(0)W >>1$, where $W$ is the electronic bandwidth 
\be
\lambda = N(0)<I^2>/(M<\omega^2>) \propto 1+N(0)W 
\ee
This is in qualitative accord with the data in Fig. (\ref{I2/w2}); the 3d, 4d and 5d metals have progressively larger band-widths and the anti-bonding part of the band has effectively a larger band-width than the bonding part.

\subsubsection{\bf Maximum $T_c$ and General Lessons.}

It is an amusing excercise to use the empirical relations to see the scale of maximum $T_c$. Such an excercise was indulged by McMillan \cite{mcmillan} and by Anderson and Cohen \cite{Cohen-Anderson} for pseudopotential metals with amusing misunderstandings resulting, for which the originators can hardly be blamed. The emphasis in these estimates was both on conditions on lattice stability and the compeition between the repulsive Coulomb pseudopotential and the attractive interaction through phonons. I give here an estimate for maximum $T_c$ based on the empirical relations discussed above, which is similar to that done about 30 years ago \cite{cmv-conf}.  This is valid for metals and compounds where a tight-binding approach to the electronic structure and electron-vibrational interaction is valid. These are the likely materials for the highest $T_c$ by e-ph.

Using the  expression, $T_c \approx <\omega> \exp {-(1+\lambda)/\lambda}$, and assuming $<\omega> \approx <\omega^2>^{1/2}$, one can use the two empirical relations to express $T_c$ in terms of $\lambda, E_c/r_0^2$ and the ion mass $M$. Now maximizing $T_c$ with respect to $\lambda$ gives that the optimal value of  $\lambda \approx 2$ and  \be
(T_c)_{max} \approx \sqrt{\frac{E_c}{2Mr_0^2}} \exp(-3/2).
\ee
One may estimate the magnitude for compounds with $Nb$ as the main element with atomic mass 92.  The typical cohesive energy per formula unit estimated from the second moment of the band is about 10 eV for Nb or $Nb_3Sn$. Taking $r_0 = 1.34 \AA$, the co-valent radius of $Nb$ gives $(T_c)_{max} \approx 30 K$. This is to be compared with the experimental value of $9.2$ K for $Nb$ and $22$ K for $Nb_3Sn$. 

The conclusion is that for the highest $T_c$ from electron-phonon interactions, one needs a small mass $M$ of the ion to give the large frequency scale of attractive interactions. In a compound, the electrons near $E_f$ must have substantial weight on the same atoms with the small $M$ so that the electron-phonon interactions and the phonon energies are related to the average stiffness $W/r_0^2$. The highest $T_c$ of the electron-phonon variety so far is $MgB_2$ which satisfies these conditions. Taking the geometric mean of the mass $\approx 13$ proton mass and $W$ from band-structure of about $10eV$ and $r_0 \approx 1 \AA$ gives a $(T_c)_{max}$ of about $60$ Kelvin while the actual value is $42$ kelvin. One cannot do better than metallic Hydrogen where the bandwidth is estimated to be about 1eV. $(T_c)_{max}$ of about $100 K$ is to be expected. These are generous limits because they exclude the reduction due to the Coulomb pseudo-potential which is expected to be especially severe for hydrogen.

 An important lesson from the study of the empirical relations in "high-temperature" superconductors of the e-ph variety and their explanations, is that the parameters determining $T_c$ are gross parameters, which depend on the average local stiffness of the lattice and on how the electronic energy of the bands changes with local deformations. Although superconductivity is a fermi-surface phenomena, the parameters determining $T_c$ are properties related to the variation of the electronic bonding energy with the local fluctuation responsible for pairing. These lessons carry over to the electronic mechanisms of pairing except that getting effective coupling constants of $O(1)$ with their main benefit - the larger high-frequency cut-offs, requires rather more stringent conditions, as we shall see below.

\section{Superconductivity from Fermion Interactions}

The theory of Fermi-liquids by Landau \cite{landau}, which was almost concurrent with the BCS theory of superconductivity \cite{bcs}, led to an enormous interest in the experimental study of the properties of liquid $He^3$ in the 1960's \cite{wheatley}. Liquid $He^3$ near the melting line is very strongly correlated with the magnetic susceptibility  about  25 times larger than a non-interacting fermi-gas of the same fermi-energy.
With BCS theory in mind, it was natural to think of pairing or superfluidity in $He^3$. Two $He^3$ atoms have a large s-wave repulsion due to the hard-core interaction. It is necessary therefore that any bound state of a pair of atoms $\psi({\bf r})$ have a node in the wave-function at the relative co-rodinate ${\bf r} =0$. The radial part of the pair wave-function must therefore vanish as $\psi({\bf r}) \propto r^{\ell}$, and so the bound-state must be in the ${\ell}$-th angular momentum. It follows from Pauli-principle that even ${\ell}$'s have total spin zero, i.e. a singlet state, while the odd ${\ell}$'s have spin $1$, i.e. a triplet state. The idea of a finite angular momentum pairing for $He^3$ therefore arose \cite{brueckner}.

Following Berk and Scrieffer's \cite{berk-schrieffer} result that ferromagnetic fluctuations are deleterious for s-wave pairing, Layzer and Fay \cite{fay-appel} suggested that such fluctuations may promote spin-triplet $\ell =1$ pairing. The discovery of such pairing in liquid $He^3$, (for reviews, see \onlinecite{leggett, vollhardt-woelfle}),
 led to further theoretical ideas and calculations. It is generally agreed that although the idea of exchange fluctuations give the correct symmetry of pairing, weak-coupling calculations based on the idea do not work quantitatively. Calculations properly taking into account the short range repulsion between the $He$ atoms and estimating interactions with constraints put  by the measured Landau parameters give the right scale of $T_c$ and its pressure dependence \cite{pfitzner-woelfle}. 

In the late 1970's superconductivity in heavy fermions was discovered \cite{steglich-ceuc2si2, ott-ube13, stewart-rmp}  and more superconducting heavy fermions continue being discovered. In these materials, the mass enhancement of the fermions is of $O(10^3)$, so that the effective fermi-energy is of the same order or smaller than the characteristic phonon energy. It was suggested that in this case the phonon attractions could not overcome the Coulomb repulsion because the concept of the Coulomb pseudo-potential is invalid and therefore the pairing must be in a finite angular momentum state \cite{cmv-hfsc}. Experiments were suggested to test this suggestion which were soon carried out \cite{hf-expts}. Transport experiments were analyzed \cite{schmitt-miyake-cmv, hirschfeld-woelfle} and they could only be understood if there was a line of zeroes of the gap-function on the fermi-surface. Such a gap function is not allowed \cite{blount, volovik-gorkov} in the spin-triplet manifold in the presence of spin-orbit scattering. Therefore $\ell=2$ pairing was to be expected. At the same time, it was known that the heavy fermion superconductors were close to anti-ferromagnetic instabilities. This led to an investigation of pairing through anti-ferromagnetic fluctuations which we have reviewed above \cite{msv} which has since been much used (and in my view abused) extensively in connection with the superconductivity in the Cuprates.
A concurrent RPA calculation \cite{scala-loh} of the pairing susceptiblity in the Hubbard model also revealed a tendency to d-wave pairing.

\subsection{Superconductivity in the dilute Fermion Gas with varying Interactions: \\Theory and Experiments for Cold Atoms}

The advent of the technique of cooling atoms in atomic traps has generated (among other things) a new class of Fermi-liquid in which the particle density $\nu$ is low but the inter-particle interactions $U$ can be varied from very weak to very strong. The {\it effective} particle-particle interaction in the low density limit is completely specified by the scattering length $a$ so that all physical properties are functions of the dimensionless coupling strength 
\be
\kappa = \frac{1}{k_fa},
\ee 
where $k_f$ is the magnitude of the fermi-vector. The theoretical deduction of the effective interaction in this situation is simpler than for other fermi-liquids. The upper cut-off of interaction energies is the Fermi-energy. The essentially exact calculations of $T_c/E_f$ possible for this problem provide a measure of the highest value to be expected from e-induced $s$-wave pairing for the complicated situations. It is not coincidental  that the maximum $T_c/\theta_D$ realized in e-ph superconductors approaches the highest value in the calculations summarized below.

The $t$-matrix approximation for effective interactions, which is exact in the low density limit, gives that  the  the scattering length $a$ is given in terms of the parameters of the Hubbard model by \cite{bec-bcs}
\be
\label{scatt-length}
\frac{m}{4\pi a} = U^{-1} + \int \frac{d{\bf k}}{(2\pi)^3} \frac{1}{2 \epsilon({\bf k})} \equiv  U^{-1}- U_*^{-1}.
\ee
Here U is the interaction parameter in the Hubbard model, $<0$ to give $s$-wave pairing, $\epsilon({\bf k}) = k^2/2m$,  and $U_*^{-1} = -\pi l_0m$; $2\pi/l_0$ is the upper cut-off in the integral over ${\bf k}$ introduced to avoid the ultraviolet divergence.

The weak-coupling limit (BCS-limit), when the attractive interaction strength is negligible compared to the kinetic energy is given by $\kappa \to -\infty$ and the opposite or Bose molecular limit by $\kappa \to  +\infty$. In between is the unitarity limit $\kappa \to 0$ where $a \to \infty$. Universal results are to be expected for physical properties for all low density attractive interaction models in these three limits. These limits are realized in experiments by tuning the interactions through the {\it Feshbach resonance}. 

\begin{figure}[ht!]
  \centerline{
  \includegraphics[width=0.7\columnwidth]{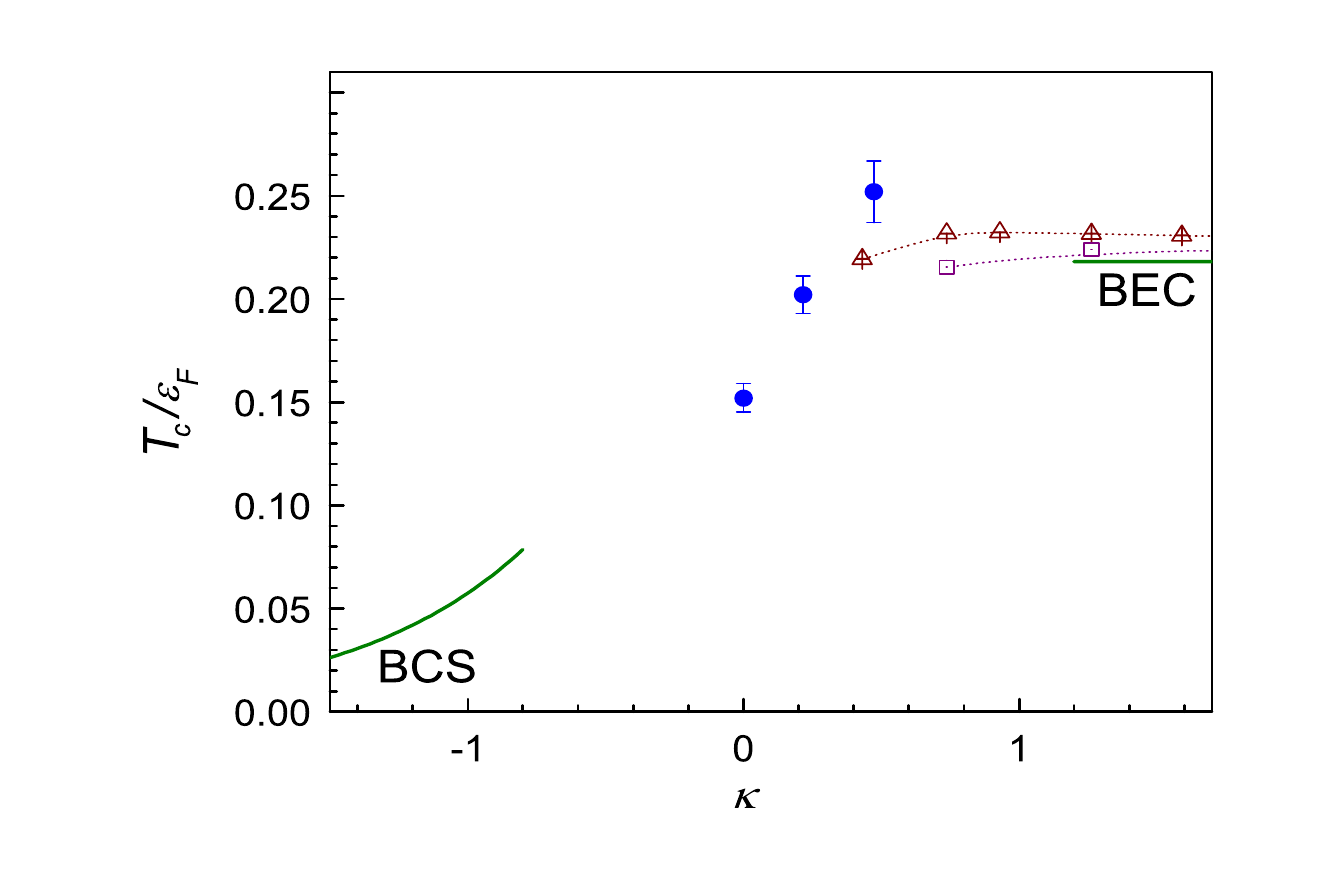}}
  \caption{Monte-Carlo calculations \cite{bec-bcs} for dilute fermion gas with varying interaction parametrized by the dimensionless number $\kappa$ around the unitarity limit, $\kappa =0$ . The results are compared to the weak interactions or BCS limit, $\kappa \to -\infty$ and to the Bose molecular formation limit, $\kappa \to \infty$. Non-universal results are obtained for the latter limit, shown by different values for different forms of potentials}\label{coldatoms/Tc/Ef}
\end{figure}

Monte-carlo methods have been used to calculate \cite{bec-bcs} $T_c/E_f$ both for interactions for free fermions around the unitarity limit  as well as for attractive interactions in the Hubbard model on the square lattice for varying density. Figure (\ref{coldatoms/Tc/Ef}) gives the results and compares them to the results in the BCS limit as well as the Bose molecular limit. The results in the unitarity limit are $T_c/E_f = 0.152(7)$. The same result is obtained in the unitarity limit for the attractive Hubbard model, as expected from  universality. Figure (\ref{coldatoms/Tc/Ef}) shows,
unexpectedly, that the results as a function of $\kappa$ are not monotonic in going from the BCS to the molecular limit, but have a maxima on the molecular side. 

The thermodynamic deductions of the properties of the cold atomic gases in the unitarity limit \cite{coldatomExpts} give a value $T_c/E_f \approx 0.2$. Given the difficulty in determining thermodynamic properties precisely as well as the non-uniformity in the density of the gas in the optical trap, this should be considered good agreement with the calculations.

\subsection{Liquid $He^3$: Pairing Symmetry, $T_c$ and Connection to Landau Theory}

\begin{figure}[ht!]
  \centerline{
  \includegraphics[width=0.5\columnwidth]{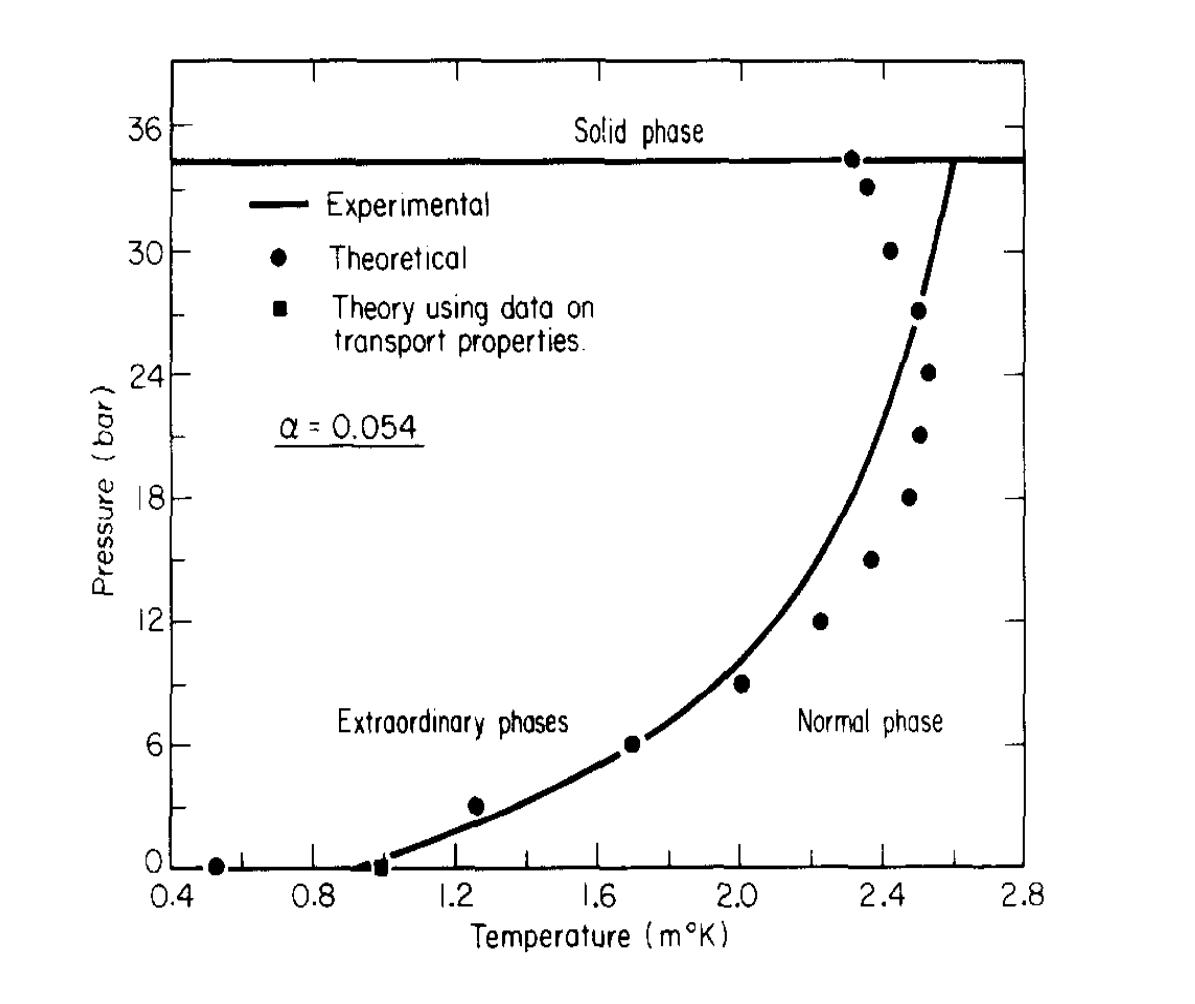}}
  \caption{$T_c$ in liquid $He^3$ as a function of Pressure compared with the calculations in Ref.(\onlinecite{PZ}) based on fitting Landau Parameters} \label{he3-Tc-P}
\end{figure}

Near the melting line of $He^3$, the effective mass is about 6 times larger, the compressibility about 15 times smaller and the magnetic susceptibility about 25 times larger than a non-interacting gas with the same density of states at the chemical potential. The transition temperature of superfluid $He^3$ might therefore be taken to represent the scale of $T_c$ to be expected for a strongly interacting fermi-liquid away from criticality.  The low energy properties are given by the Landau theory and thermodynamic and transport properties have been measured extensively.  $T_c/E_f$ is about $10^{-3}$. Much is to be learnt from the work on this problem, not the least is why $T_c$ is so low. Variation of $T_c$ with pressure together with results of an interesting calculation by Patton and Zarnalingham \cite{PZ} are shown in fig. (\ref{he3-Tc-P}). 

The general pairing vertex has been given in Eq. (\ref{BSEqn.}) and Fig.(\ref{pairingvertex}) in terms of the irreducible particle-particle vertex $I_S$. The relation of $I_S({\bf k,k+q, k'})$ to the Landau parameters is interesting to explore. With ${\bf k,-k}$ and ${\bf k+q, k-q}$, all close to the fermi-surface,  $I_S({\bf k,k+q})$ is specified by $I_S(\theta,\phi)$, where  $\theta$ is the angle between ${\bf  k}_F$ and ${\bf  k+q}_F$ and  $\phi$ is the the angle between the planes formed by ${\bf k, k+q}$ and ${\bf k', k'-q}$.  For pairing, we need only $I_S(\pi,\phi)$. 
The forward scattering limit  ${\bf q} \to 0, \omega \to 0$ of $I$ are the province of the Landau theory of Fermi-liquids. 
The "A" Landau interaction parameters are given by the forward scattering limit, $(v_f q)/\omega \to 0$, then   $\omega \to 0$ of $N(0) I_S(\theta,\phi) = A_S(\pi,\phi)$. They are decomposed into different angular momentum channels:
\be
A_S^{\ell} = A_S(\pi,\phi)P_{\ell}(\phi).
\ee
The forward scattering sum rule (imposed by Pauli-principle) requires 
\be
\sum_{s,\ell} A_{s,\ell}=0.
\ee
 Patton and Zarnalingham \cite{PZ} argued that since the irreducible pairing vertex needs to be calculated in the limit $\omega/(v_Fq) <<1$, it should be related to the  
 "F" Landau parameters, which are in related to the irreducible vertex in  the other limit, $\omega/(v_Fq)\to 0$ and then $q \to 0$. "F's" are related to the
"A's" through
\be
F_{s,\ell} = A_{s,\ell}/(1+A_{s,\ell})
\ee
The measured compressibility and the spin-susceptibility at various pressures provide $F_{0,0}$ and $F_{1,0}$ respectively as a function of pressure, while the specific heat provides $F_{0,1}$. The Landau parameters are expected to rapidly decay with increasing $\ell$ due to the short-range nature of the effective interactions; if one assumes that it is saturated by $\ell =0$ and $\ell=1$, $F_{1,1}$ may be extracted using the forward scattering sum-rule. 

 Patton and Zarnalingham \cite{PZ}  found  that the pairing interactions extracted with the assumptions above give repulsion in the $\ell=0, s=0$ and attraction in the $\ell=1, s=1$ channel. They could get a surprisingly good value for $T_c (P)$ over the whole range of pressures (see fig. (\ref{he3-Tc-P})) using the BCS formula $T_c \approx \omega_c \exp{(-1/\lambda)}$ if they use that the upper cut-off $\omega_c(P) \approx \frac{1}{20}E_f(P)$. 
 
The weak part in the argument is the assumption that $\omega/V_fq <<1$ may be used with $q \to 0$ to related the vertex to the "F's" and the lack of any explanation of the value of the cut-off.  Detailed calculations of the Landau parameters and the extension of Landau theory to vertices with finite momentum transfer (to remove one of the the weak parts of the argument) have since been done using partly phenomenological and partly microscopic approaches \cite{vollhardt-woelfle}. The complete such calculations done by Pfitzner and Woelfle \cite{pfitzner-woelfle} carefully using the exact equations for the vertices and using empirical information on the measured thermodynamic and transport properties of the normal state of liquid $He^3$ to constrain limiting values of the solutions and interpolating to get the complete vertex. This is the best kind of microscopic theory suitable for such difficult problems. With these calculations the singlet $\ell =0,2$ channels are shown to be repulsive and the triplet $\ell =1$ channel is attractive with the right variation of coupling constants with pressure to provide the variation of $T_c$. 
 
One of the important lessons to be learnt from such studies is that the effective coupling constants are determined by the $A_{\ell}$ parameters (and their finite momentum extensions which are expected to be smoothly varying) and their modulus is always smaller than 1. This represents the physical fact that the dimensionless coupling constants involve the product of the density of states and the interactions. Large renormalizations always (nearly) cancel out in this product because of cancellations of self-energy and vertex renormalizations. In Landau theory these cancellations are related to conservation laws. The other lesson is the empirical lesson that $\omega_c$ is an order of magnitude smaller than $E_f$. This cannot be obtained from Landau theory but any reasonable microscopic calculation gives that the cut-off of the fluctuations goes down as the F parameters increase. For example the sound-velocity goes down as the compressibility or $F_0$ increases, the spin-fluctuation frequencies go down as the magnetic susceptibility or $(m^*/m)(1/1+F_0^a)$ increases, etc.

Let us turn briefly to the microscopic calculations from weak-coupling theory. In the paramagnon model, the best of these is due to Levin and Valls \cite{levin-valls}. The local Hubbard interaction 
\be
H_{int} = I \int d{\bf r}d{\bf r}' n({\bf r})n({\bf r}') \delta ({\bf r}-{\bf r}')
\ee
promotes ferromagnetic exchange in free-fermions (no lattice potential) and associated increase of the amplitude of  ferromagnetic spin-fluctuations for $I < I_c$, the critical value for ferromagnetism. This model does not describe the physics of $He^3$ very well because of the substantial range of the hard-core interactions compared to inter-particle spacing; it does not give the right values of the Landau parameters or their pressure dependence. But nevertheless several features of our interest in relation to calculations of $T_c$, from fluctuations induced by particle-hole interactions, from such a model (and its modifications) have substantial educational value because RPA respects conservation laws. 
Levin and Valls performed such calculations in the Hubbard model to calculate both the Landau parameters as well as solve the Eliashberg equations with the fluctuation spectra obtained in the calculations. The important results of their analysis and calculations are:

(1) The pairing interaction in the $\ell=1$ channel $V_1$ has  a direct dependence on $I$ but so does the effective mass as well as a  $m^*/m$. The renormalized pairing interaction parameter $\lambda_1$depends on the  product $V_1(m^*/m)^{-1}$.  Moreover the self-energy correction in the $T_c$ formula depends on parameter $\lambda_0$. This goes up with $I$. This reduces  $T_c$ and may be taken to contribute to an effective reduction of cut-off if one insists on using the BCS formula for $T_c$ even for finite $\ell$.

(2) The net effect still is that $T_c$ goes up with increasing $I$ except close to the ferromagnetic transition, where it swings downwards towards $0$. 
This is  due to the pile up of the fluctuation spectra to low energies for $I$ close to $I_c$. This has two deleterious effects on $T_c$:  it increases inelastic scattering and reduces the cut-off $\omega_c$. 

These conclusions are consistent with those from the analysis of the Eliashberg equations with a general form of pairing and self-energy kernel discussed above as well as with the Ward identities. A Gutzwiller-type variational wave-function with a Hubbard model on a lattice \cite{vollhardt-woelfle} with less than 1/2 filling gives much better account of the ground state properties of normal $He^3$ than does a paramagnon model. But the extensions to the excitation spectra with the same basic physics can not work, since putting the model on a lattice promotes antiferomagnetic correlations and does not lead to pairing interactions in the $\ell =1$, triplet channel. 

\subsection{Experimental results on Heavy Fermions}

Heavy Fermion compounds show superconductivity, generally associated with an AFM quantum-critical point, (but note the interesting case of $CeCu_2Si_2$ under pressure, where two forms of criticality, AFM, and Mixed-Valence each seem to have an associated superconducting region \cite{holmes-miyake}), although the converse is not true; AFM quantum critical points in some heavy-fermion compounds are not accomapanied by superconductivity. Superconductivity always appears to be in a finite angular momentum state and is not due to electron-phonon interactions. 
\begin{figure}[ht!]
  \centerline{
  \includegraphics[width=0.7\columnwidth]{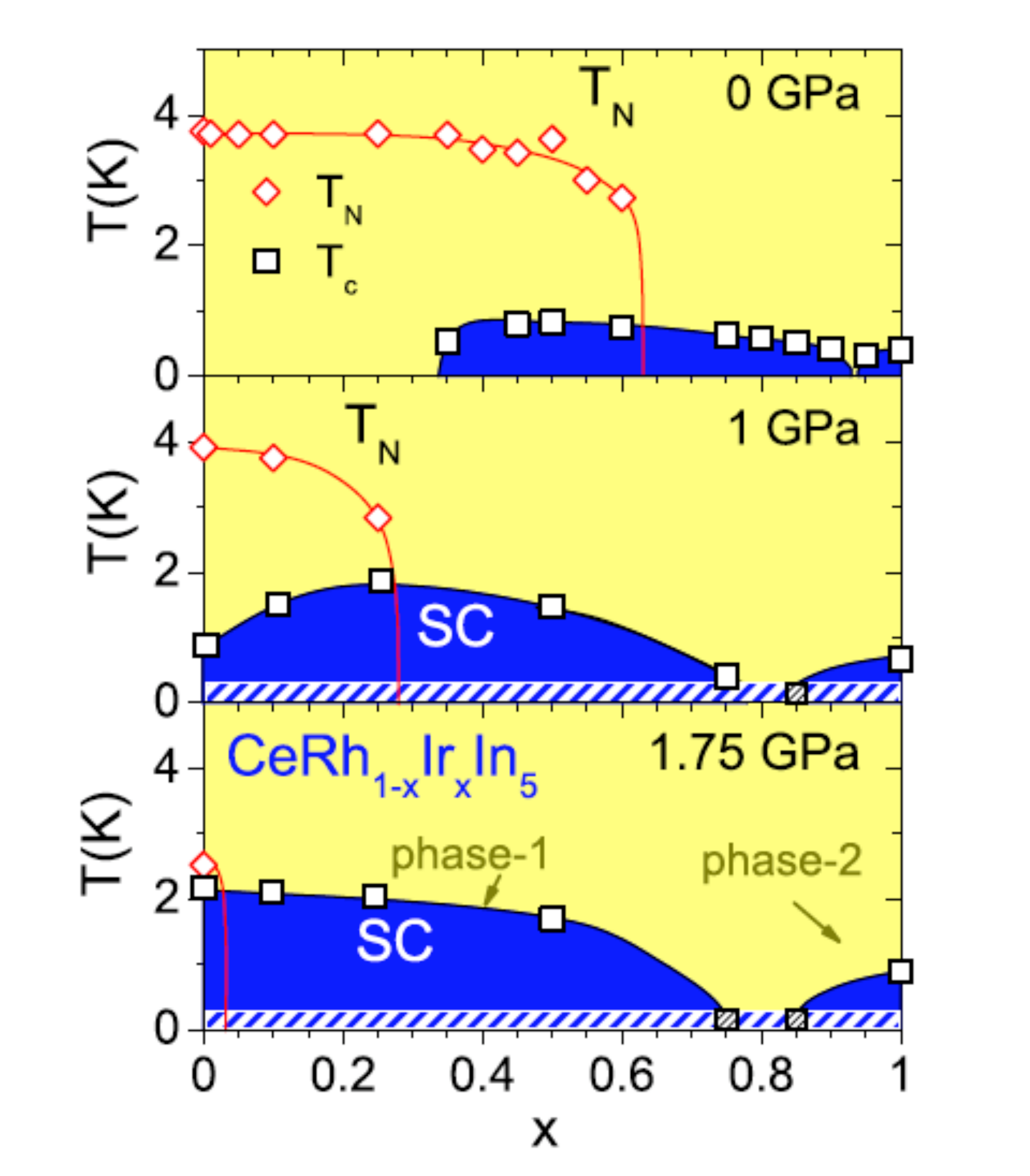}}
  \caption{Phase Diagram of a 115 compound and alloys at various pressures shown. Figure is reproduced from Ref.(\onlinecite{thompson-hf}).} \label{115-phaseDiag}
\end{figure}

Heavy Fermions in rare-earth and actinide compounds are the ultimate realization of the ideas of analyticity and continuity which underlie the Landau quasi-particle idea. In their fermi-liquid regime, the effective mass enhancement in several heavy-fermion compounds is $0(10^3)$ and the quasi-particle renormalization residue $z$ is $O(10^{-3})$. This situation changes in the quantum-critical regime where the quasi-particle idea breaks down and transport and thermodynamic properties are not those of a Fermi liquid. This is a beautiful example of how $z \to 0$ as $T\to 0$, only logarithmically, produces completely different physical properties at low temperatures than $z \approx 10^{-3}$. 

Knowing the fermi-liquid renormalizations is not as useful to deduce parameters for superconductivity in heavy-fermions as in liquid $He^3$ for two reasons:  superconductivity is near qcp's, where such renormalizations are singular and the qcp are at large $q$-vectors, where Landau parameters are not defined. However, the energy scales are set by the renormalizations given by the Landau parameters above and are therefore essential to bear in mind. They are ideal systems to study magnetic fluctuations by inelastic neutron scattering. But only in a few of them are such results available near quantum-criticality because for the technique to be fully effective requires large single-crystals. 
 \begin{figure}
  \centerline{
  \includegraphics[width=0.7\columnwidth,bb=0 0 500 500]{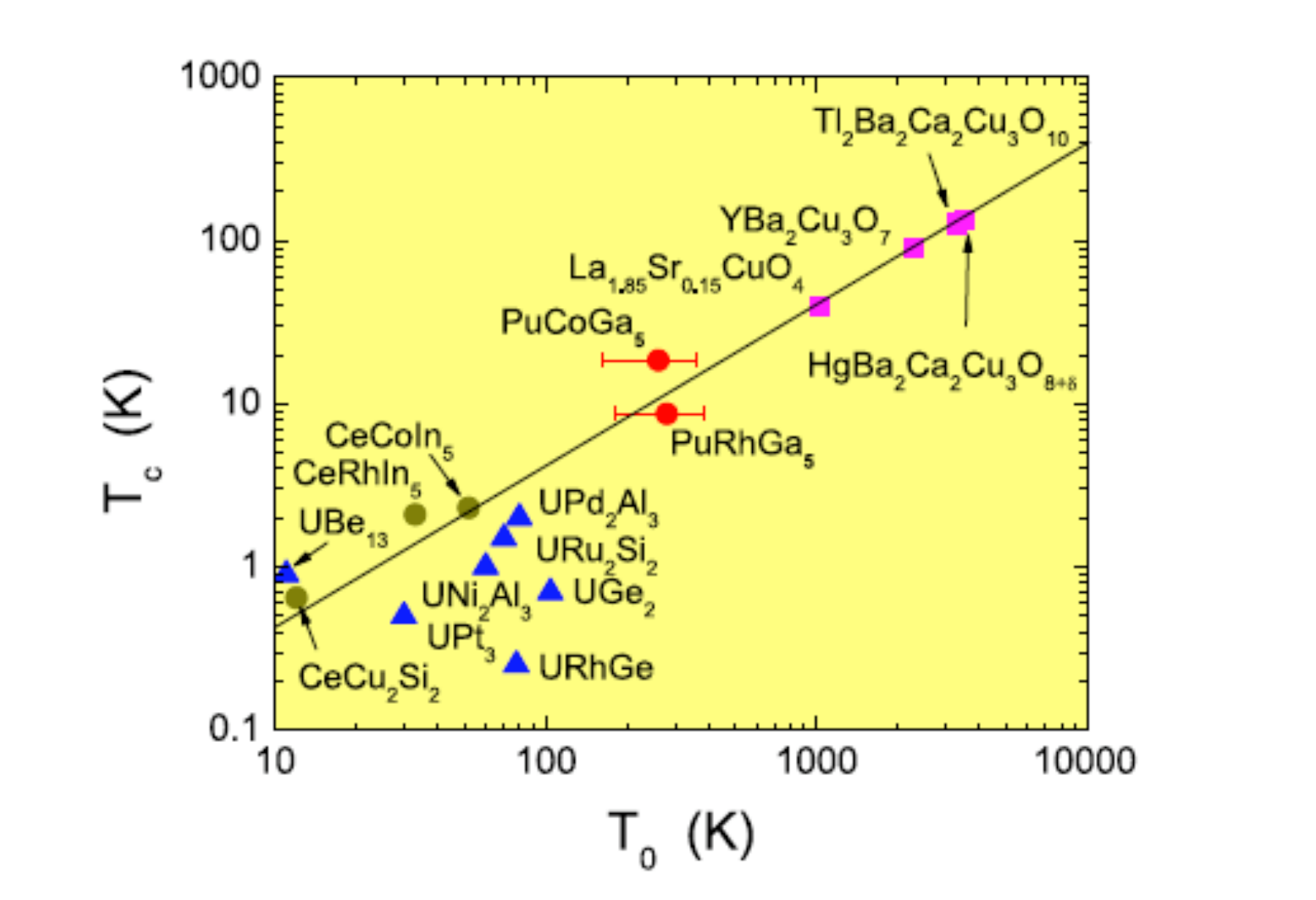}}
  \caption{$T_c$ vs. $T_0$, a measure of the Fermi-energy obtained from magnetic susceptibility measurements from several heavy-fermion superconductors as well as Cuprates; figure is reproduced from Ref.(\onlinecite{thompson-hf}).} 
 \label{Tc/To}
\end{figure}
From the study of the thermodynamic and  transport properties (such as residual resistivity, temperature dependence of resistivity, ultrasonic attenuation, thermal conductivity and nuclear relaxation rates) in the fermi-liquid regime, the leading renormalizations in heavy fermions were found to be qualitatively different from that in liquid $He^3$. The renormalizations are characteristic of a sub-set of fermi-liquids in which the single-paricle self-energy is very weakly dependent on momentum compared to on energy. This idea gives that \cite{cmv-hf}
\bea
\label{hfrenorm}
m^*/m = 1/z = (1+ F_0^s) \\ \nonumber
\kappa/\kappa^0 = (m^*/m)/(1+F_0^s) = O(1) \\ \nonumber
\chi/\chi_0 = (m^*/m)/(1+F_0^a) = O(1).
\eea
The enhancement of the specific heat of $O(10^3)$ gives $z$ of $O(10^{-3})$; similar enhancement of susceptibility gives $F_0^a$ of $O(1)$; the lack of renormalizations in ultrasonic attenuation and zero-temperature resistivity give that $F_0^s \approx 1/z$. The temperature dependence of the resistivity in the Fermi-liquid regime is $\propto T^2$ with its coefficient renormalized by $O(1/z^2)$ which also follows from the momentum independence of the self-energy.
These ideas are also realized in microscopic calculations based on dynamical mean-field theory \cite{dmft-voll, dmft}, which also start with the assumption that the self-energy is momentum independent. 

At a microscopic level, it is understood that the renormalizations in the heavy-fermion lattice are closely connected with the Kondo effect of the magnetic ions in the bath of conduction electrons. This physics is well-incorporated in the models of the lattice by a variety of different techniques, the most versatile of which is the dynamical mean-field theory which yields results similar to Eqs. (\ref{hfrenorm}) together with the frequency dependence of the one-particle spectra. A theory of  two interacting Kondo impurities \cite{jones-cmv-wilkins} with interactions among the magnetic ions competing with the Kondo effect at either also exists. The theory yields a low energy effective Hamiltonian for the problem with several interesting terms, including pairing interactions among the quasi-particles. 
A low energy Hamiltonian for the lattice at a similar level has not yet been derived but is essential for a detailed theory of superconductivity in them. As discussed above, one can motivate the d-wave symmetry of superconductivity in most of the heavy-fermions through exchange of antiferromagnetic fluctuations. But this does not answer the question of the energy scales of fluctuations in the critical regime. All we know is just that the renormalizations discussed through Eqs.(\ref{hfrenorm}) give the scale of the upper limit of the cut-off energy of the pairing fluctuations to be down by $z$ from the bare parameters, i.e. of $O(E_f)$, the renormalized fermi-energy. (As discussed already, the value of z serves to renormalize the cutoff scale as well as the coupling constants in general accord with Landau fermi-liquid theory.) The actual experimental values of $T_c$ are down from this by only between one and two orders of magnitude from the upper cut-off, see Fig. (\ref{Tc/To}). One may speculate that this is associated with the fact that unlike liquid $He^3$, superconductivity in heavy-fermions generally always occurs in the region around quantum-criticality, see Fig.(\ref{115-phaseDiag}), and the criticality is not of the conventional variety where the energy scale of fluctuations $\to 0$ as criticality is approached. 

In the one case in which quantum-critical fluctuations have been studied in single-crystals by neutron scattering the criticality is  of the topological or local variety \cite{schroder} discussed in Sec. II: the fluctuation spectra scales as $\omega/T$ but is essentially q-independent; the space and time are not tied together as in Gaussian quantum-criticality. This particular compound is however not superconducting. But, based on the measured specific heat, and resistivity measurements, one can be fairly sure that criticality is of the same variety in the heavy fermions where superconductivity occurs  with relatively large values of $T_c/E_f$. The measured specific heat $\propto T log T$ and temperature dependence of resistivity $\propto T$ cannot be understood in three-dimensional materials by the Gaussian variety of quantum criticality \cite{rosch-rmp}.

The attempts to derive quantum-criticality in heavy fermion lattice \cite{si} are equivalent to the self-consistent solution of the single-impurity Kondo problem in a conduction electron bath which is close to antiferromagnetic quantum-criticality \cite{maebashi}. The breakdown of the Kondo effect near magnetic singularities is doubtless an important aspect of the problem but such methods do not represent a solution of the criticality problem. The issue of quantum-criticality in heavy fermions is a beautiful largely unsolved problem; it is also not possible to discuss their "high" $T_c$ without understanding the fluctuation spectra near quantum criticality and its coupling to fermions. At a phenomenological level also, one is limited in this discussion due to lack of measurements of the fluctuation spectrum or single-particle spectra in the heavy-fermions near quantum-criticality. Such measurements will require quality and size of samples similar to some of the Cuprates.

\subsection{Superconductivity in the Cuprates}
 
  \begin{figure}[tbh]
  \centering
\includegraphics[width=0.6\textwidth]{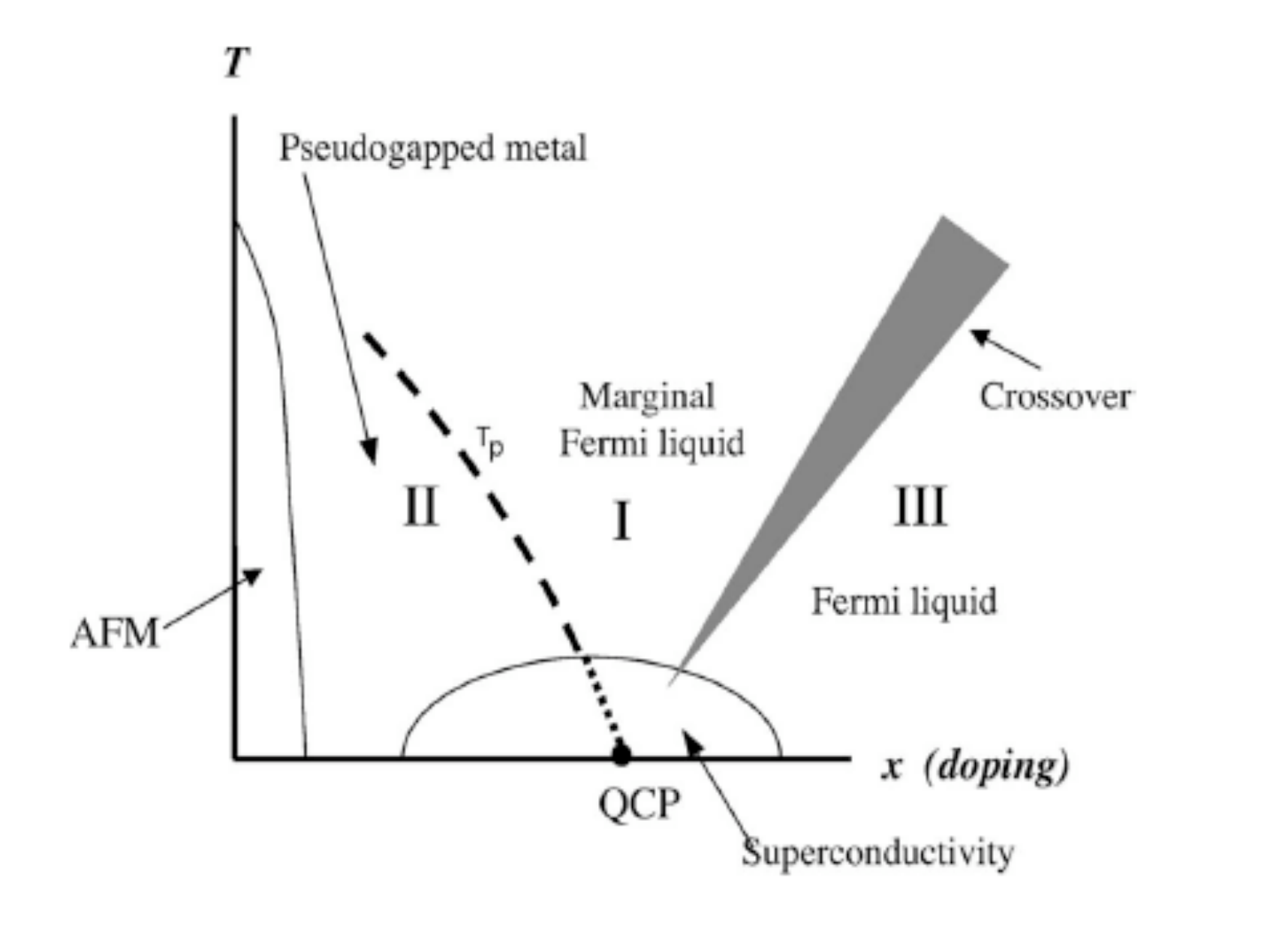}
\caption{The Universal Phase diagram for hole-doped cuprates based on properties which show characteristic changes across the lines drawn in all Cuprates. The demarcation region between Region II and the AFM region has not been determined clearly by experiments or theory.}
\label{CupratePhase}
\end{figure}

 The family of Cuprates have the highest transition temperatures discovered so far. This phase diagram of Fig.(\ref{CupratePhase}) organizes the extensive experimental investigations in the Cuprates in the last twenty years. It is the universal phase diagram in the sense that the lines drawn demarcate change in properties in the same manner in every Cuprate family investigated. (There are of-course Cuprates with special properties in one or the other regions of the phase diagram which are not shared by other Cuprates.) The phase-diagram is meant to be schematic and correct only in its topology. 
 
There have been innumerable ideas and calculations to understand the properties of the Cuprates. I will only summarize the work which seeks to understand all the principal normal state properties as well as superconductivity with one set of ideas and has predicted unique results which have been tested by experiments.

The central organizing feature of the phase diagram is a quantum-critical point from which two lines emanate which separate the normal state in to three parts; The region of superconductivity is located around the qcp. The location of the qcp within the diagram varies with respect to the region of $T_c \ne 0$ in different cuprates; the maximum $T_c$ is generally not at the doping $x_c$ of the qcp. In Region I, quantum-critical properties are observed in Resistivity, optical conductivity, Raman Scattering, nuclear relaxation rate on Cu-nuclei (but not O-nuclei which show fermi-liquid type relaxation rates), tunneling into a conventional metal and single-particle spectra measured by angle-resolved photoemission (ARPES). The superconductive pairing is in the "d-wave" channel. The great mystery has been the nature of Region I and of Region II. Region II,  the so-call pseudogap region abuts the boundary to Region I whose properties are determined by quantum-critical fluctuations (qcf). 
Through extensive experimental investigations, it is now generally believed that Region II represents a phase which competes with superconductivity, much the same as antiferromagnetism competes with superconductivity in the heavy fermions. Needless to say, Region II has no long-range anti-ferromagnetic order and near the region of maximum $T_c$, in at least one of the heavily investigated Cuprates, the magnetic correlation length is about a lattice constant, which is the scale of magnetic correlations in non-interacting fermions.

Given the general proposition that the pairing symmetry and $T_c$ are the properties of the normal state just above $T_c$, the central problem for superconductivity in the Cuprates is the understanding of the nature of the quantum-critical fluctuations, which is a problem which can be usefully attempted only if the nature of Region II is understood since given the phase diagram, the fluctuations in Region I are the quantum-fluctuations of the order parameter in Region II. There is physics in both regions which we had not come across before. For example, as we will see, the fluctuations in Region I are quite unlike the Gaussian fluctuations discussed in Sec. II C. The symmetry breaking in Region II remains controversial.

There have been two valid reasons for skepticism that Region II has a distinct order parameter. One is that it  the extensive investigations on the Cuprates with every available tool and some newly invented had not discovered any credible broken symmetry and the second that although the specific  heat decreases near below the line demarcating Region I and II, there is no evidence for any singularity in it as in the typical classical phase transition. However, in the last five years, an elusive proposed order \cite{cmv-prb97, simon-cmv1, cmv-prb06} has indeed been discovered experimentally in four distinct families of Cuprates \cite{kaminski, fauque, greven, mook}. Moreover the model representing the observed classical phase transition to this order has a non- diverging specific heat at the transition \cite{baxter}, whose quantitative value is consistent with the experiments \cite{gronsleth}.

\begin{figure}[tbh]
\centering
\includegraphics[width=0.5\textwidth]{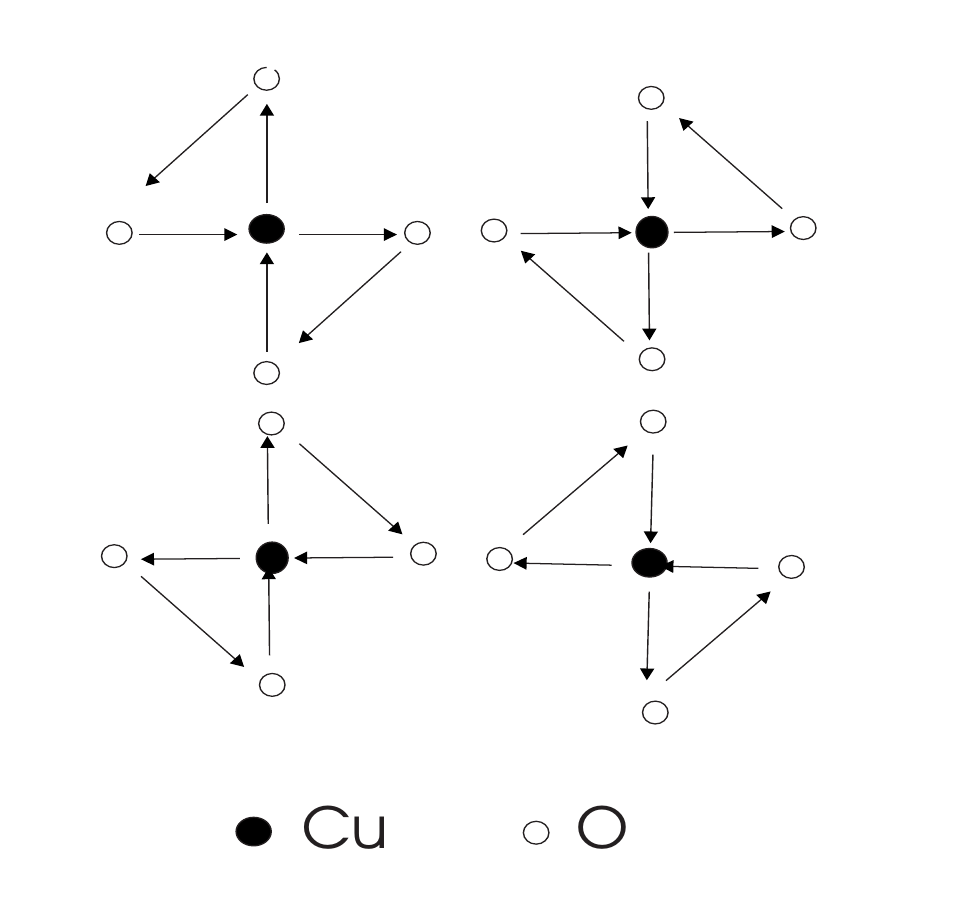}
\caption{The four possible domains with the symmetry of the observed order in the underdoped Cuprates.}
\label{loop-order}
\end{figure}

\begin{figure}[tbh]
\centering
\includegraphics[width=0.7\textwidth]{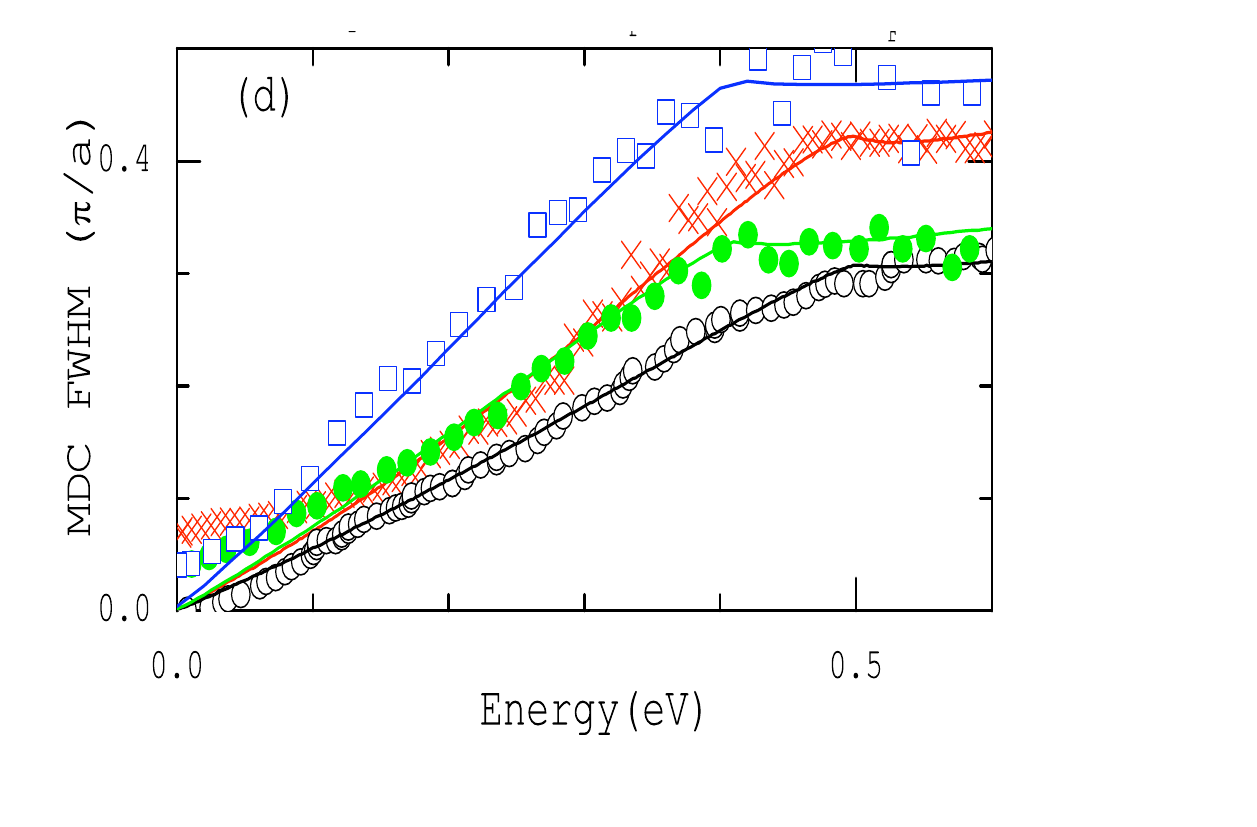}
\caption{The linewidths of the Momentum Distribution curves for different cuprates as a function of the energy. The imaginary part of the self-energy is obtained by multiplying this linewidth with the bare fermi-velocity. The detailed references for each cuprate are given in Ref.(\onlinecite{lijun-prl08})}
 \label{mdcwidth}
\end{figure}

\begin{figure}[tbh]
  \centering
\includegraphics[width=0.8\textwidth]{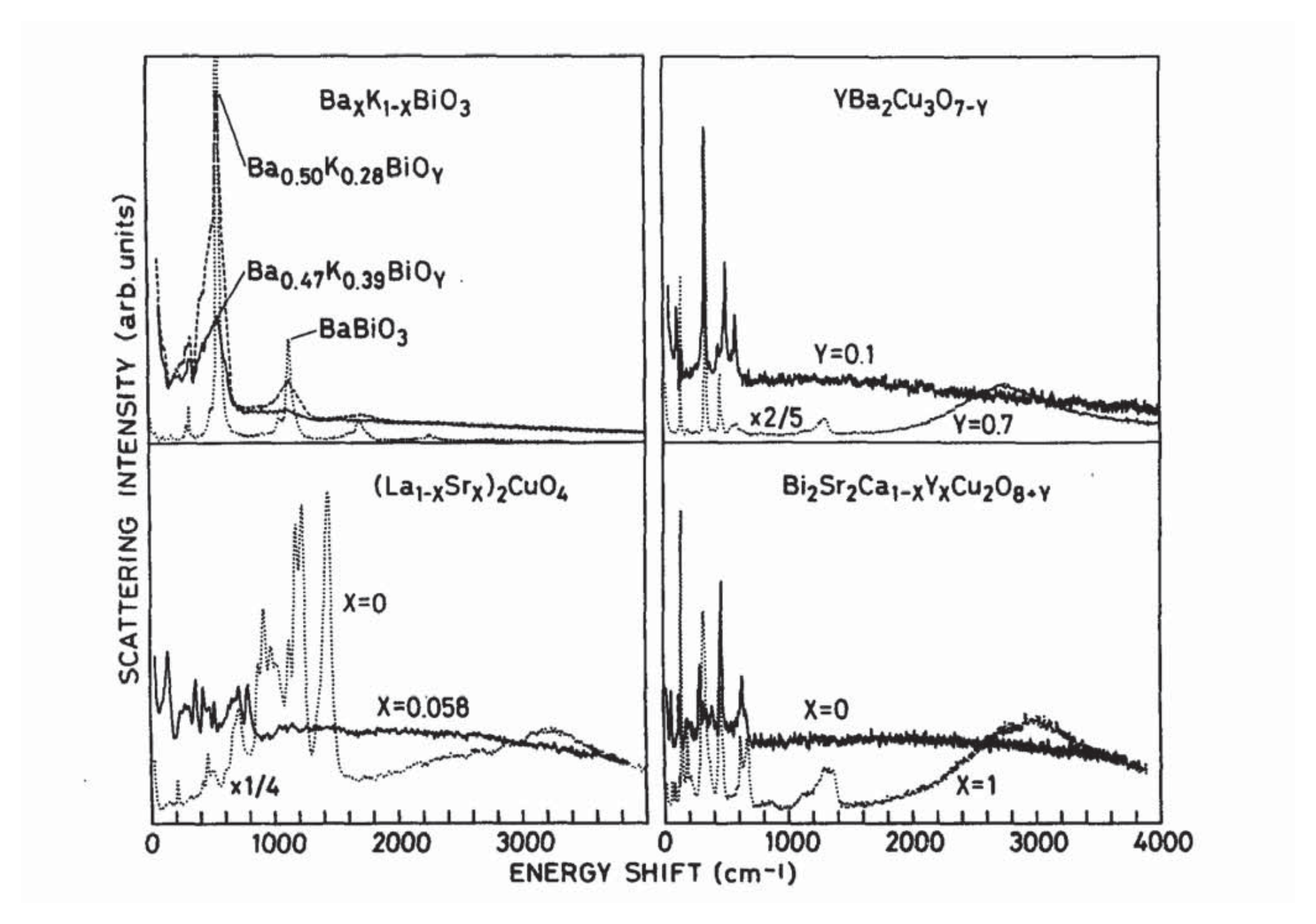}
\caption{Universality of Raman scattering in Cuprates near and $Ba_xK_{1-x}BiO_3$ for compositions of doping with the highest $T_c$. The figure is taken from (\onlinecite{sugai-raman}).}
\label{sugai-raman}
\end{figure}

I will now summarize the theoretical results briefly. In a two-dimensional Cuprate model with three degrees of freedom per unit-cell, representing the $d_{x^2-y^2}$-orbital of Copper and the $p_x$ and the $p_y$-orbitals on the Oxygen atoms on the x-direction and the y-direction respectively of the Copper, together with on-site and nearest neighbor interactions,  two phases which are odd under time-reversal due to spontaneously generated orbital loop-currents but preserving the translational symmetry were shown to be locally stable in a mean-field calculation. The phase observed has the symmetry of one of these phases but there are some other details \cite{kaminski, fauque, greven, mook} which are different and obtainable only in a more complicated model \cite{weber-prl09}. The observed as well as the predicted phase has a symmetry which can be characterized as the ordering of a polar time-reversal odd vector ${\bf L}$. ${\bf L}$ describes the loop-current pattern shown in Fig.(\ref{loop-order}) which preserves even-ness only in reflection on one of the four-possible symmetry planes of the square lattice. ${\bf L}$ has four-possible orientations, corresponding to the four possible domains of the ordered phase.
In the quantum-fluctuation regime, the instantaneous pattern of currents is described by vectors ${\bf L}_i$, which vary in direction among the four-possible orientations. The quantum-fluctuations are described  by the correlation function \cite{aji-cmv-qcf, aji-cmv-qcf-pr} $<{\bf U}^+_i(t) {\bf U}_j(t')>$, where ${\bf U}_i$ is the generator of the rotation among the four-configurations of ${\bf L}_i$. It turns out that in the fluctuation regime the discreteness of ${\bf L}_i$ can be replaced by a continuous vector so that the fluctuations are related to the fluctuations of a quantum-fluctuations of a $2-$dimensional rotor. In a model including dissipation due to the coupling of the fluctuations of  ${\bf U}_i$ to the fluctuations of the fermion-current, the spectral-function $\chi"({\bf q}, \omega)$ of the fluctuations in the quantum-critical regime has been derived (\cite{aji-cmv-qcf}, \cite{aji-cmv-qcf-pr}) to be of the form
\begin{eqnarray}
\label{eq:flucspec}
\Im\chi({\bf q},\omega) &=& \begin{cases}
 -\chi_0 \tanh(\omega/2T), &|\omega| \lesssim \omega_c;  \\
0,  &|\omega| \gtrsim \omega_c.
\end{cases}
\end{eqnarray}
This is precisely of the form which was suggested \cite{mfl} to account for the singular transport properties in Region I of the phase diagram and is the basis of the marginal fermi-liquid. The single-particle scattering rate from such fluctuations can be calculated, the imaginary part of the self-energy for frequencies much larger than the temperature is given simply by
\begin{eqnarray}
\label{selfenergy}
\Im\Sigma(\omega, {\bf k}) &=& -\frac{\pi}{2} \lambda({\bf k})\begin{cases}
 |\omega|, & |\omega| \lesssim \omega_c  \\
 \omega_c, & |\omega| \gtrsim \omega_c.
\end{cases}
\end{eqnarray}
Here $\lambda({\bf k})$ is a coupling function whose derivation is discussed below.  Fig. (\ref{mdcwidth}) shows that in the $(\pi,\pi)-$ direction, this prediction is fulfilled in all Cuprates in which ARPES measurements have been carried out. The scattering rate at any energy $\omega$ is essentially $\propto \sum_{\bf q} \int _0^{\omega} \chi"({\bf q}, \omega')$. Therefore the linearity in $\omega$ of the scattering rate up to some energy and constancy thereafter is a direct proof of the fluctuation spectra given by Eq. (\ref{eq:flucspec}). Earlier experiments with greater resolution at lower energies providing evidence the crossover from linearity in $T$ to linearity in $\omega$ have been reviewed \cite{abrahams-cmv-pnas}.  Fig. (\ref{mdcwidth}) is also a proof that a distinct fluctuation spectra with a sharp cut-off $\omega_c \approx 0.5 eV$ exists universally in the cuprates. These critical fluctuations themselves for $q \to 0$ are directly 
observed in Raman scattering (but the experiments have not been carried out all the way to the cut-off energy), see Fig.(\ref{sugai-raman}), where evidence for the universality is presented through $S(\omega) = (1+n(\omega/T))\chi"(\omega)$ in the limit $q \to 0$.

The spectra of Eq.(\ref{eq:flucspec}) is quite unlike the Gaussian critical spectra discussed in Sec.IIC.  The singularity at $(\omega, T) \to 0$ does not affect the bulk of the spectra at all which extends at all $T$ to 
$\omega_c$, which as we will infer from experiments is about $E_f/4$. Second and most curiously, the critical spectra has no spatial scale, the concept of a dynamical critical exponent $z$ is lost. 

Given that these singular fluctuations determine the properties above $T_c$ including the scattering rate of the fermions, it is natural to ask if they promote superconductive pairing with the observed d-wave symmetry and with the right order of magnitude of $T_c$. As has already been noted in Sec. II, the spectra of the fluctuations ({eq:flucspec}) is ideal for high $T_c$ on the basis of Eliashberg theory.  It has a high upper cut-off, and it has the least inelastic scattering possible in a quantum-critical spectra. However, the q-independence of the spectra makes one wonder if it can promote d-wave pairing. To investigate this, the momentum dependence of the coupling of the fermions to the fluctuations has been calculated.

The coupling  of the fermions to the fluctuations has been calculated \cite{shehter-aji-cmv}. In the continuum limit, ${\bf U}({\bf r})$ is the angular momentum operator generating rotations of ${\bf L}({\bf r})$.
Therefore it can only couple to the local angular momentum of fermions. So the coupling is of the form
\be
\label{coupl}
H_{int} \propto \int d{\bf r} \sum_{\sigma}g \psi^+({\bf r},\sigma) ({\bf \hat{r} \times \hat{p}}) \psi({\bf r},\sigma) {\bf U}({\bf r}) + H.C.
\ee 
$H_{int}$ has also been derived for the fermions in a two-dimensional model of the Cuprate lattice and the coupling constant $g$ estimated in terms of the same microscopic model which gives the symmetry of the observed order and its approximate magnitude. 
It is useful to note that Eq. (\ref{coupl}) is the natural orbital angular momentum analog of the familiar collective spin-fluctuation coupling to to spin-flip excitations of fermions.

We may write Eq.(\ref{coupl}) in momentum space;
\be
\label{coup-kspace}
H_{int} = \sum_{{\bf k, k'}, \sigma} g  i ({\bf \hat{k}} \times {\bf\hat{k'}})  \psi^+({\bf k},\sigma)\psi^+({\bf k'},\sigma){\bf U}({\bf k-k'}) + H.C.
\ee
The coupling constant of the scattering of fermions to the fluctuation spectrum can be extracted from the ARPES data in the normal state, Fig. (\ref{mdcwidth}) \cite{lijun-prl08}. From such measurements, one deduces that the coupling constant $\lambda_0$ for all Cuprates measured by ARPES is between about $0.7 $and $1$ and the cut-off $\omega_c$ is between $0.4$ eV and $0.5$ eV.
The lattice generalization of Eq. (\ref{coup-kspace}) also predicts that the scattering rate varies $\propto a + b\cos(4\theta)$ where $\theta$ is measured from the $\pi,\pi$ direction with a variation of  about a factor of 2 going from the $\pi,\pi$ to the $\pi,0$ direction. 
 
 The momentum-dependence of coupling, even though the spectrum itself is momentum dependent is crucial to the symmetry of superconductivity promoted by the critical fluctuations. This is seen as follows:
 Integrating over the fluctuations in Eq.(\ref{coup-kspace})gives an effective vertex for scattering of fermion-pairs:
\begin{align}
\label{hpair}
H_{pairing} \approx & \sum_{{\bf k}\sigma{\bf k'}\sigma'} \Lambda({\bf k},{\bf k}')
c^{\dagger}_{\sigma'}(-{\bf k}')c^{\dagger}_{\sigma}({\bf k}')c_{\sigma}({\bf k})c_{\sigma'}(-{\bf k}); \notag \\
\Lambda({\bf k},{\bf k}') = & \gamma(k,k')\gamma(-k,-k') \Re\chi(\omega=\epsilon_{\bf k}-\epsilon_{\bf k}').
\end{align}
This is exact to $O(\lambda \frac{\omega_c}{E_f})$, where $\lambda$'s are the dimensionless coupling constants exhibited below. 
 In the  continuum approximation for fermions near the fermi-energy, $\gamma({\bf k}, {\bf k}') \propto  i({\bf k} \times {\bf k}')$. The pairing vertex is then
\begin{equation} \label{kxk'}
\Lambda\left(\textbf{k},\textbf{k},\right) \propto -({\bf k} \times {\bf k}')^2\Re\chi({\bf{ k}}-{\bf{k'}},\omega).
\end{equation}
Since  $\Re \chi({\bf k}-{\bf k'}),\omega) < 0$  for $-\omega_c <\omega < \omega_c$, independent of momentum, the pairing symmetry is given simply by expressing $({\bf k} \times {\bf
k}') ^2$ in separable form~:
\begin{align} ({\bf k} \times {\bf k}') ^2 &= 1/2 \left[(k_x^2+k_y^2)(k_x^{'2}+k_y^{'2})
-  (k_x^2-k_y^2)(k_x^{'2}-k_y^{'2})\right.\nonumber \\
&- \left. 4(k_xk_y) (k'_x k'_y)\right].
\end{align}
Pairing interaction in  the $s$-wave channel is repulsive, that in the two $d$-wave channels is equally attractive, and in the odd-parity channels is zero. The factor $i$ in $\gamma({\bf k}, {\bf k}')$, present because the coupling is to fluctuations of 
time-reversal odd operators, is crucial in determining the sign of the interactions of the pairing vertex.

To estimate $T_c$, we use what has been discussed in Sec.(II) about the effect of inelastic scattering on finite angular momentum pairing. $T_c$ is given approximately by
\be
\label{Tc}
T_c \approx \omega_c\exp(-(1+|\lambda_s|)/|\lambda_d|),
\ee
 where $\lambda_s$ is the
coupling constant which appears in the normal self-energy and $\lambda_d$, the coupling constant which appears in the pairing self-energy. From the measurements summarized above and Eq.(\ref{kxk'}), $\lambda_d/\lambda_s \approx 1/2$. Using the deduced value of $\lambda_s$ and $\omega_c$ from the ARPES measurements, one estimates a value of $T_c \approx 100 K$. 

Although $T_c$ is expected to reduce in the underdoped region due to the competing phase and in the overdoped region due to the change in the spectra to an incoherent spectra below a cross-over scale, no 
quantitative calculations for these effects exist.

\subsection{The case of the Fe-Pnictides}

 The newly discovered superconductivity in the Fe-pnictides is also quite unlikely to be induced through interaction with lattice vibrations. A recent review is Ref.(\onlinecite{pnictide-review}). The highest $T_c$ in this class of compounds so far is about $50 K$ in $RFeAs(O_{1-x}F_{x}): (R=Ce, Pr, Sm, Nd,$ etc). The "parent compound" at $x=0$ is metallic but antiferromagnetic.  The lack of significant observable feature in the specific heat at the high superconducting transition temperatures raises doubt as to whether bulk superconductivity in this structure of the Fe-Pnictides has indeed been found. There is also some evidence that this structure may have a two-phase co-existence as a function of doping.  A closely related new structure of Fe-Pnictides called
122 appears to form good single crystals. Thermodynamic data indicates bulk superconductivity. The phase diagram of this class of materials appears similar for both hole doping and electron doping.
 \begin{figure}[tbh]
  \centering
\includegraphics[width=0.8\textwidth]{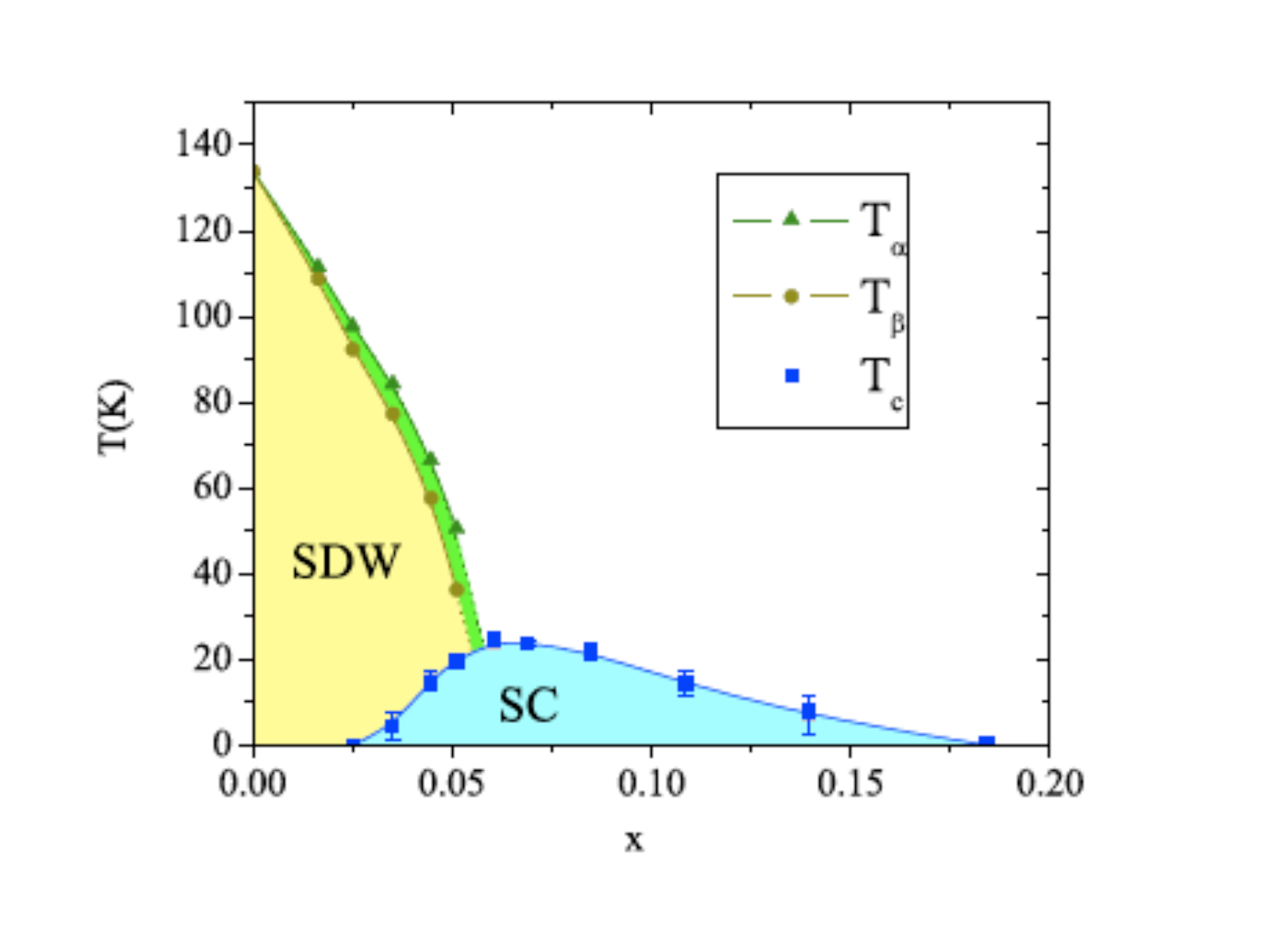}
\caption{The Phase diagram of Co-doped FeAs2 \cite{fisher-pnictide-phdia}, \cite{canfield}. The green triangles mark the transition to an altered structure while the black circles mark the antiferromagnetic transition. Superconducting region is shown in blue. }
\label{FeAsPhase}
\end{figure}
The superconducting region is organized around a quantum critical point, see Fig.(\ref{FeAsPhase})  where an AFM/structural transition temperature $\to 0$ with change in doping. The mystery of the Cuprates: the nature of the ordered phase on one side of the critical point is absent. Moreover, the anisotropy in resistivity of these compound is less than an order of magnitude - they are properly considered three-dimensional. The fermi-surface has five sheets, most prominently a pair of electron-pockets centered at the zone-center and a pair of electron pockets at the zone-faces. At this point, the symmetry of superconductivity is not unambiguously known. ARPES experiments indicate that there is a gap everywhere on the Fermi-surface. This may well be an extended s-wave form of pairing, as has been suggested \cite{mazin, kuroki, bangchoi}. As discussed earlier, AFM fluctuations would favor such a symmetry of pairing for an appropriate fermi-surface; the form of the fermi-surface in this class of compounds is the right kind.
Given the phase diagram, it is also reasonable to infer that the AFM/structural quantum-criticality is responsible for the high transition temperatures. But as discussed above, the Gaussian quantum-criticality is bad for $T_c$, both due to the prefactor and the inelastic scattering, especially for the case of pairing not of the simple s-wave variety.  The important question therefore is the nature of the quantum-critical fluctuations. Only a limited set of experiments are at present available on good single crystals to answer this question and no conclusive statements can be made yet.

 \begin{figure}[tbh]
  \centering
\includegraphics[width=0.8\textwidth]{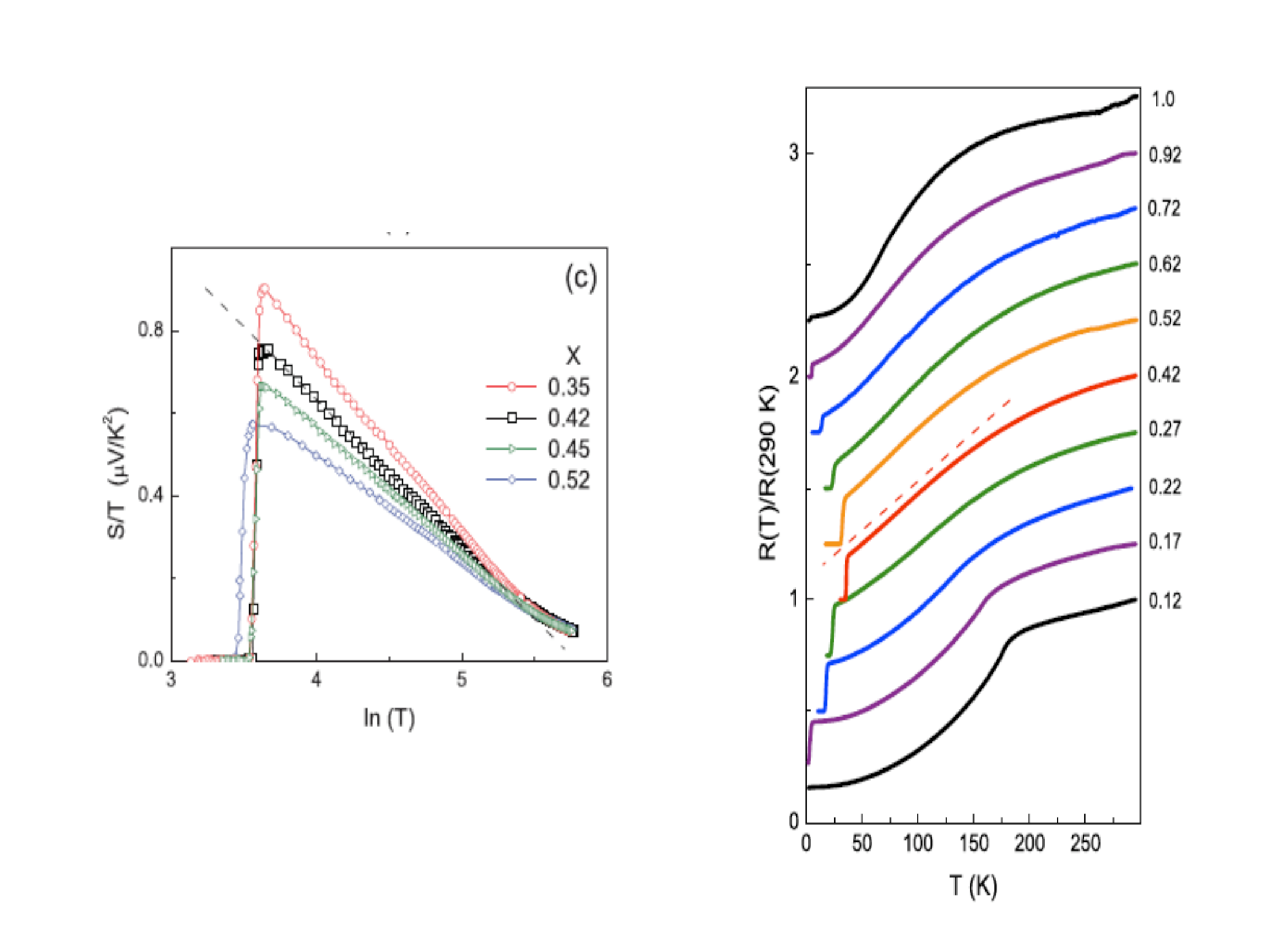}
\caption{On the left: Thermopower \cite{gooch} in K-doped FeAs2 indicating a possible $T\ln T$ electronic entropy near the doping for highest $T_c$, indicating possible quantum-criticality; On the right: Resistivity \cite{fisher-pnictide-phdia} in Co-doped FeAs2 for various concentration, showing a possible linear in T resistivity near the doping of highest $T_c$, indicating scattering from a singular fluctuations spectra consistent with that which gives $T\ln T$ for the electronic entropy. Both are charaterisitic of Marginal Fermi-liquids.}
\label{thermo-resis-pnictides}
\end{figure}

Two experiments suggest that quantum-criticality may be in the same universality class as the Cuprates, i.e. the fluctuations have a weak low frequency singularity and a broad nearly constant distribution in frequency. One is the measurement of thermopower for hole doped compound to deduce the electronic contribution to entropy in the normal state, see fig. (\ref {thermo-resis-pnictides}) and the other is the resistivity measurements, also shown in the same figure but for the hole-doped compound. The thermopower measurements are consistent with an entropy $\propto T \ln (T)$ down to $T_c$ from an upper cut-off of about the room temperature in a region close to the highest $T_c$. Similarly, the resistivity is linear in temperature down to $T_c$ in a similar region. Neither of these properties are characteristic of fluctuations around a Gaussian quantum critical point, especially in 3d systems \cite{rosch-rmp}.

More experiments are needed to investigate whether the Fe-Pnictides have the same class of quantum-criticality as the Cuprates and the heavy-fermions. Experiments which would be helpful are measurements of the Raman and inelastic neutron scattering spectra, measurements of single-particle scattering rates by ARPES measurements and of transport scattering rate through optical conductivity measurements. Also analysis of ARPES spectra in the superconducting and normal state following such experiments  to decipher the spectrum of the pairing glue will undoubtedly be forthcoming. Inelastic scattering in the normal state to indicate the variation of the spectra with $(q,\omega)$ to see whether the bulk of it has very slow q-dependence and a high energy cut-off as well as $\omega/T$ scaling would be especially helpful.

On the theoretical issues, several models were proposed which are misguided multi-orbital generalizations of models for Cuprates. The opinion is converging to models similar to those with Hund's rule couplings and local repulsions of magnitude similar to the band-width. This is the class of models relevant to itinerant magnetism \cite{herring-it-mag} in metals like Ni or Fe. For such models, spin-polarized variational band-structure models 
give the right values for ground state magnetization and fermi-surfaces. Solution of such models for their fluctuation spectra is quite another matter. The physics of itinerant ant-iferromagnetism and ferromagnetism in such situation are not fermi-surface effects, the spin-gaps are more than an order of magnitude larger than the transition temperatures and essentially the entire band is affected. Possible large spin-gaps in the AFM pncitides should be looked for in spin-polarized photoemission. This always means that critical modes are soft over a large range in q-space leading to small intrinsic correlation lengths and huge reductions of transition temperatures due to the large entropy of the collective modes. There are no really good theories of this even for the classical fluctuations above finite temperature phase transitions. How this goes over to zero-temperature transitions where the soft modes may become scale-invariant over the entire momentum range is a marvelous unsolved question. Understanding of such quantum-criticality may be the key to understanding superconductivity in such systems.

 \subsection{The case of $Ba_{1-x}K_xBiO_3$}
 I wish to single out the special case of $Ba_{1-x}K_xBiO_3$ (and $BaBi_{1-x}Pb_xO_3$) because they have $s$-wave superconductivity which is most likely driven by e-e processes. This as explained below may be the key to superconductivity with much higher transition temperatures. 
 
\begin{figure}[tbh]
  \centering
\includegraphics[width=0.6\textwidth]{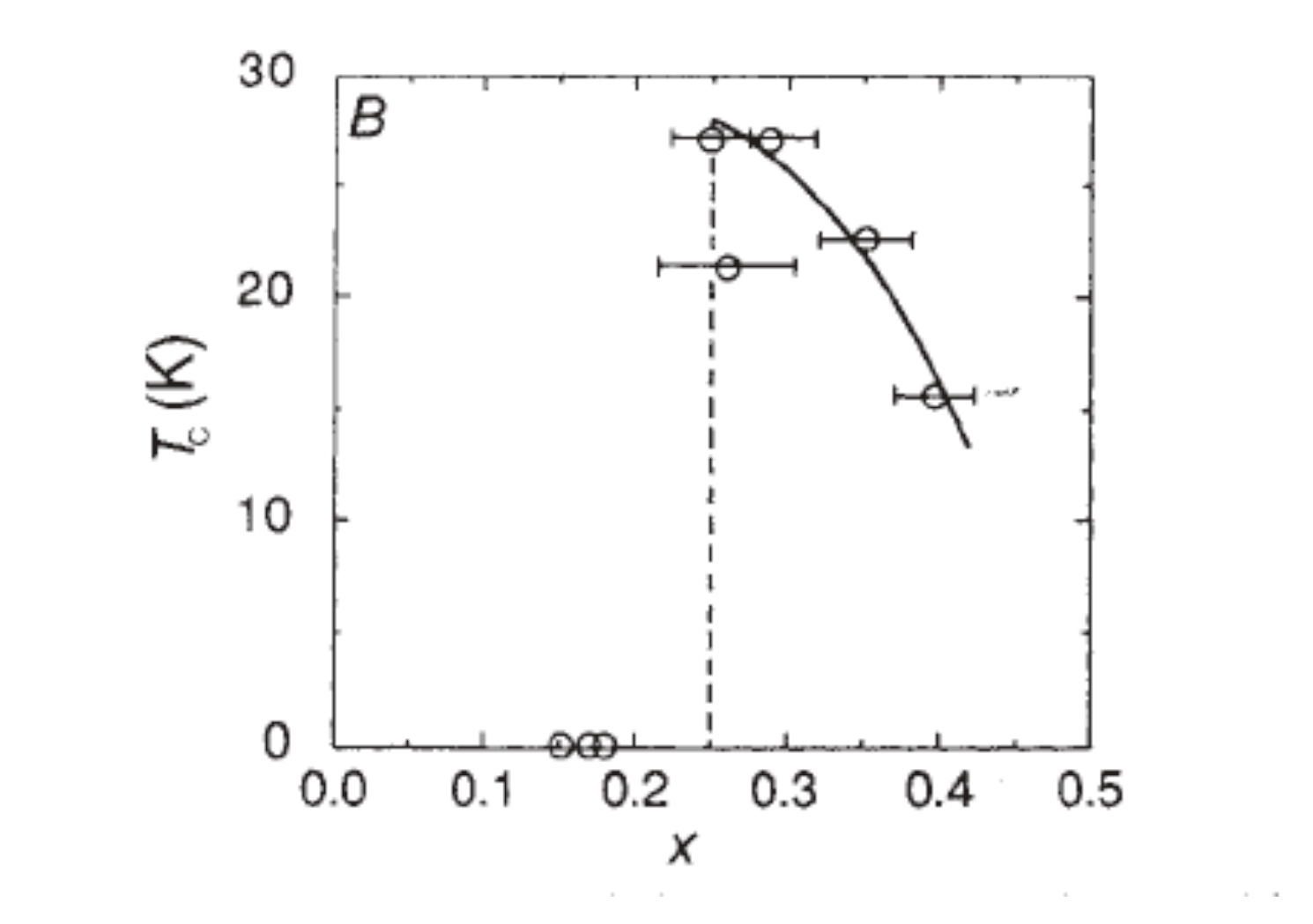}
\caption{$T_c$ vs. $x$ in $Ba_{1-x}K_xBiO_3$; the region for small $x$ has incommensurate structural order with inequivalent sites for the nominally $Bi^{3+}$ and $Bi^{5+}$. The diagram is taken from Ref.(\onlinecite{jorgensen})}
\label{babi-phdia}
\end{figure}

As seen in fig. (\ref{Tc-gamma}), the $T_c$ of $Ba_{1-x}K_xBiO_3$ is an order of magnitude larger than that of other metals of similar low electronic density of states at the chemical potential. It was suggested \cite{cmv-negU} that the superconductivity could not be due primarily to electron-phonon interactions and some detailed calculations \cite{savrasov-babio3} support this. $Bi$ is one of about a dozen elements in the periodic table which skip a valence in any of the compounds they form; the formal valence of $Bi$ is always $3^+$ or $5^+$, except for the anomaly of $BaBiO_3$, where it is $4^+$. In the cubic structure, this compound has 1 electron per unit-cell but  the crystal structure is distorted producing a diamagnetic insulator with two inequivalent sites for $Bi$, one with an oxygen co-ordination favoring $Bi^{3+}$ and the other $Bi^{5+}$. On doping with $K$, this distortion  disappears in what appears to be a first-order fashion to a metallic phase with a high transition temperature. The phase diagram is shown in fig.(\ref{babi-phdia}).
 
 The proper definition of an effective local interaction energy $U(x^+)$ of an ion at a valence $x^+$ is the difference of the ionization and the affinity energy, i.e.
 \be
 U(x^+) \equiv E((x+1)^+) + E((x-1)^+) - 2 E(x^+).
 \ee
 Therefore, for example the positive affinity energy of spin-1/2 configuration $Cu{(2+)}$ to spin-0 configuration $Cu{(1+)}$ and the positive ionization energy to the spin-1 configuration $Cu{(3+)}$ is equivalent to a repulsion $U$ so that magnetic states are favored. Correspondingly, ions like $Bi{(4^+)}$, which have a negative affinity and a negative ionization energy may be modelled by an attractive local interaction $U$.   
When the average valence is the skipped valence and the ions fluctuate quantum-mechanically between the two valences on either side, the fluctuations are that of a local Cooper pair. Superconductivity requires the further process of phase coherence between such fluctuating pairs at different sites induced by the kinetic energy.

The chemistry and physics leading to this "negative" $U$ is not well understood. It doubtless depends on the fact that the valence on either side of the skipped valence forms closed shells so that $U$ calculated from the gas-phase ionization and affinity energy is much lower than for adjacent elements \cite{cmv-negU}, although still positive. This is likely to be reduced in a low energy model by polarization processes and by  nearest neighbor repulsions \cite{cmv-negU}. But a quantitative understanding of why in the solid state, irrespective of their chemical surroundings, such elements always skip valence needs further investigation.  

 The phase diagram of the form in fig. (\ref{babi-phdia}) and some of its other properties were obtained in a model with negative $U$ and repulsive nearest neighbor interactions \cite{cmv-negU}. A detailed investigation on this model has been carried out \cite{taraphdar-negU}. I wish to highlight only the experimental features which  suggest that this compound has unusual properties, besides the high $T_c$ indicative of pairing due to electron-electron interactions. 
 A limitation on our knowledge of this compound is that the extrapolated residual resistivity of the samples measured suggests a mean-free path due to impurities of only a few lattice constants. I hope better samples of the compound can be made and its properties in the pure limit studied more thoroughly.
 
\begin{figure}[tbh]
  \centering
\includegraphics[width=0.8\textwidth]{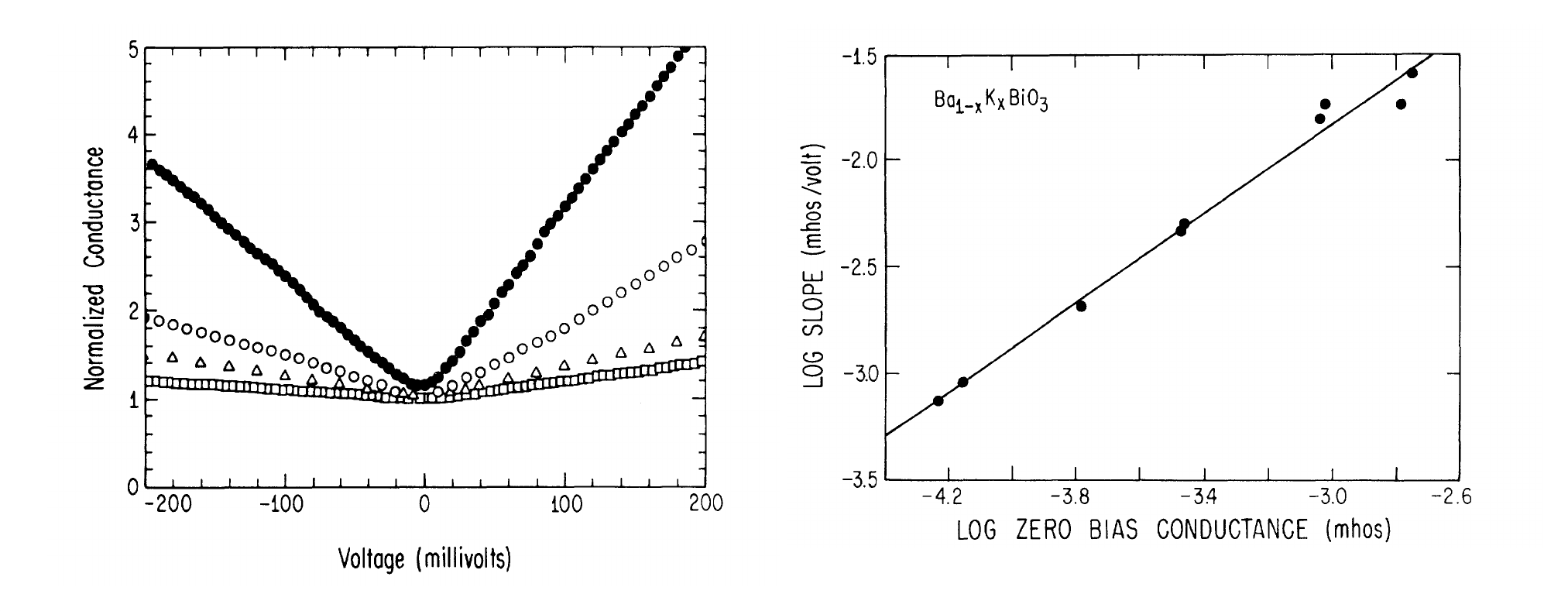}
\caption{Tunneling data from Sharifi et al. \cite{dynes-babio3} and connection of the slope of conductance to Tc}
\label{babitunn}
\end{figure}

 In the insulating state at low doping, transport and optical gaps differ by nearly an order of magnitude. In $BaBiO_3$, the transport gap is 0.24 eV while the optical gap is about 2 eV \cite{{taraphdar-negU}}. This remarkable deviation has been well explained in the model of local electronic attraction \cite{taraphdar-negU} by the formation of Cooper pairs in the insulating state. The ac conductivity in the metallic state deviates from the Drude form and resembles in some experiments the form in the Cuprates. Tunneling measurements \cite{dynes-babio3} show a conductance 
 $G(V) \propto |V|$ for $|V| >> T$. This is characteristic of a single particle scattering rate $\propto |\omega|$, the energy measured from the chemical potential. This is of the same form as observed through ARPES in Cuprates near 'optimal' doping, as well as with the peculiarity in the optical conductivity. What is even more interesting is that a systematic study of the slope $G(V)/|V|$ as a function of doping shows a correspondence with the measured $T_c$. These results are shown in fig. (\ref{babitunn}). This suggests, unsurprisingly, that the fluctuations responsible for the anomalous normal state scattering are also responsible for pairing.

 There is evidence for anomalous fluctuations in Raman scattering, shown in Fig. (\ref{sugai-raman}). They have the same form as the critical continuum observed in the Cuprates. However, the $|\omega|$ dependence of the single-particle scattering rate and in transport requires that such a critical continuum exist over a large momentum region. There are no experiments to look for this nor is there any theory which leads from a negative $|U|$ model to such critical fluctuations. 
 While more experiments in better samples are called for, a tentative case can be made on the basis of the Raman scattering and the tunneling data that the superconductivity in $Ba_{1-x}K_xBiO_3$ is also due to  the new universality class of fluctuations.
 
 The importance of this compound is that it has isotropic singlet superconductivity. From the considerations  earlier, this is the class of electronically induced superconductivity which has the best chance of having the highest $T_c$, because (1) the cut-off scale of the fluctuations is an electronic energy scale and (2) because the effects of inelastic scattering and self-energy are least deleterious for $T_c$ in the $s$-wave channel. But $T_c$ in $Ba_{1-x}K_xBiO_3$ is unimpressive because of its exceptionally low electronic density.
 
 That $Bi$ is not unique in regard to skipping valence is shown experimentally by the example of $Tl$ doped $PbTe$. Here at very low doping evidence \cite{fisher-chKondo} of fluctuations between the two valences, $Tl^{1+}$ and $Tl^{3+}$, i.e. of Cooper-pair fluctuation or  charge Kondo effect \cite{schmalian-chargekondo}, has been presented. This turns into a superconducting state with further doping as the kinetic energy increases. $T_c$ rises to about $1.5 K$ at about $1.5 \%$ doping. Unfortunately, the compound cannot be doped further.
  
Fabrication of compounds with valence skipping ions listed in ref. (\onlinecite{cmv-negU}) is suggested so as to stabilize the skipped valence as the average valence, hopefully with large electronic density. Unfortunately, they generally seem to be unstable towards one or the other side of the skipped valence; so special fabrication techniques may be required. This appears a likely route to high $T_c$ with the added benefit of being in compounds which are more or less isotropic.

 \section{ Appendix: Relation Among Parameters Determining \\
 $\lambda$ in e-ph Interactions}
 
 {\bf 1. "Constancy" of $N(0)<I^2>$}
 
 Consider the Hamiltonian for the non-degenerate band of tight-binding orbitals 
\be 
H_0 = \sum_{i<j, \sigma} t_{ij} a_{i\sigma}^+a_{j\sigma} = \sum_{{\bf k},\sigma} \epsilon_{{\bf k}}a_{{\bf k}\sigma}^+a_{{\bf k}\sigma};~~~\epsilon_{{\bf k}} = n t \cos({\bf k}.{\bf a}).
\ee
A representation of the electron-phonon interaction which very efficiently includes the local field effects is a model in which the tight binding orbital moves rigidly with the ions. The electron-phonon interactions are then obtained by simply modulation of the transfer integral:
\be
t_{ij}({\bf r}) \approx t_{ij}({\bf a}) + \frac{\partial t_{ij}}{\partial {\bf r}}\cdot ({\bf u}_j-{\bf u}_i)  +...~~,
\ee 
where ${\bf u}_i$ is the displacment of the ion at site $i$ from equilibrium. Moreover 
\be
 \frac{\partial t}{\partial {\bf r}} \approx - \frac{t}{r_0}\frac{\bf a}{|{\bf a}|},
 \ee
 where $r_0$ is the characteristic radius of the tight binding orbital.
 This yields an elecron-phonon Hamiltonian
 \be
 \label{h-e-ph}
 H_{int} = \sum_{{\bf k, q}, \nu}\sqrt{\hbar/(2M\omega_{{\bf q} \nu})}  I_{\bf k, q}^{\nu} a_{{\bf k+q}\sigma}^+a_{{\bf k}\sigma}(b_{{\bf q} \nu} + b_{{\bf q} \nu}^+),
 \ee
 where
\be
\label{I(k,q)}
 I_{\bf k, q}^{\nu} = i \frac{{\bf a}_{\alpha}\cdot \varepsilon_{{\bf q}}}{|a| r_0} \big(v_{{\bf k} \alpha} - v_{{\bf k+q} \alpha}\big).
 \ee
Here $v_{{\bf k} \alpha}$ is the velocity, $ \hbar^{-1} \partial \epsilon_{{\bf k}}/\partial k_{\alpha}$.

The parameter $<I^2>$ occurring in $\lambda$ is the average of the coupling function $ I_{\bf k, q}^{\nu}$ over the fermi-surface:
\be
<I^2> = \frac{\int dS_{{\bf k}}\int dS_{{\bf k}'} \sum_{\nu} | I_{\bf k, q}^{\nu}|^2 v_{{\bf k}}^{-1}v_{{\bf k}'}^{-1}}{\int dS_{{\bf k}}\int dS_{{\bf k}'}v_{{\bf k}}^{-1}v_{{\bf k}'}^{-1}}.
\ee
With (\ref{I(k,q)}), this is easily evaluated to give \cite{friedel2}
\be
\label{noI2}
N(0)<I^2> \approx (1/r_0^2) \sum _{{\bf k} < {\bf k}_F} \epsilon_{{\bf k}} = E_c^0/r_0^2
\ee
where $E_c^0$ is the cohesive energy. This relation was used \cite{friedel2} to understand the near-constancy of $N(0)<I^2>$ in transition metals and their alloys.

A  modification of this calculation \cite{cmv-eph} to include the first correction due to the non-orthogonality $S$ of the nearest neighbor tight binding orbitals gives that
\be
N(0)<I^2> \approx (1\pm S)E_c^0/r_0^2,
\ee
where the $(\pm)$ sign holds if the fermi-level falls in the antibonding/bonding part of the band. This reflects merely that the velocity in the antibonding/bonding parts of the band is larger/smaller approximately by $(1\pm S)$.

 {\bf 2. Relation of average electron-phonon scattering 
 to average lattice stiffness}
 
 We wish to derive \cite{cmv-conf} the observed relation shown in Fig.(\ref{I2w2} between $<I^2>$ and the average lattice stiffness $M<\omega^2>$. The phonon frequencies $\omega_q$ are renormalized from their {\it rigid-ion} values $\Omega_q$ due to the creation of particle-hole pairs created by the lattice deformation:
 \be
 \label{omega}
 \omega_q^2 \approx \Omega_q^2 + 2 \omega_q \Pi (q,0)
 \ee
 where $\Pi (q,0)$ is the self-energy
 \be
 \Pi (q,0) = \sum_k |g_{k,k+q}|^2 \chi(k,k+q)
 \ee
 and $\chi(k,k+q)$ is the electronic polarizability at zero frequency. 
 
 The average rigid-ion stiffness is given by the second-derivative of the cohesive energy  of the lattice:
  \be
 M<\Omega^2> = \sum_{\alpha}\frac{\partial ^2 E}{\partial R_{\alpha}^2}|_{R=R_0}
 \ee
 Contributions to $E(R)$ come from (1) The sum of the one-electron (band-structure energy) discussed above and (2) core-core repulsion energy minus the electron-electron repulsion energy, the latter having been counted twice in a self-consistent band-structure energy. For any long-wavelength property (2) sums to zero and it is negligible for consideration of the average stiffness compared to that of (1) because it is slowly varying with inter-site separation while (1) varies on the scale of the variation of the overlap integrals. In fact chemists calcualting the structure and vibration frequency of metals have found that good results are obtained by neglecting it altogether. For transition metals and compounds, we may note further the justification that about $60-80\%$ of the cohesive energy is provided by d-electrons alone. Even their crystal structure is correctly predicted on the basis of (1) alone.
 
 Taking the second derivative of $E(R)$ 
   \be
 M<\Omega^2> \approx  (1\pm S)E_c^0/r_0^2 \approx N(0)<I^2>.
 \ee
Consider the second term in (\ref{omega}).  If $U_{k,k+q}$ is the residual electron-electron interaction between two tight-binding orbitals,
 \be
 \chi(k,k+q) = \chi^0(k.k+q)/(1+U_{k,k+q}\chi^0(k,k+q))
 \ee
 where $\chi^0(k,k+q)$ is the zero order electronic polarizability at zero frequency,
 \be
 \chi^0(k,k') =-(1/\pi^2)\int_{-W/2}^{W/2} d\epsilon \int_{-W/2}^{W/2} d\epsilon'\frac{f(\epsilon)-f(\epsilon')}{\epsilon-\epsilon'} Im G(k, \epsilon)Im G(k', \epsilon')
 \ee
 Using $(1/\pi) Im G(k, \epsilon) =\delta( \epsilon-\epsilon_{\bf k})$ and ignoring logarithmic corrections, this for typical $k,k'$is just $N(0)^2 W$. Therefore 
using Eq. (\ref{h-e-ph}, \ref{I(k,q)} ) for $g(k,k')$ gives
 \be
 2\sum_{{\bf k,q}}M\omega_{\bf q} |g_{k,k+q}|^2 \chi^0({\bf k,k+q}) \approx - N^2(0)W <I^2>
 \ee
For most interesting cases $N(0)W \gtrsim1$, so that we must consider the change of phonon frequencies due to the electronic polarizability and local field corrections.  For large values of $k,k+q$, which alone are of interest, we may take the effective interaction $U_{k,k+q} \approx W$, as one finds in the t-matrix approximation.  
This estimate in Eqs.(\ref{noI2}, \ref{omega}) gives
\bea
N(0)<I^2>/M<\omega^2> &\approx & (1+N(0)W)
 \eea   
This derivation is necessarily crude. It works qualitatively (and semi-quantitiatively with an overall correction factor of about 1/4) because we are calculating average of a quantity over the entire zone. These results are discussed in relation to Fig.(\ref{I2/w2}) earlier. The basic physical points used, that local field corrections or modulation of bonds due to lattice vibrations are essential in understanding the details of e-ph interactions and phonon dispersion and that the parameters determining $T_c$ are inter-related, is well borne out in detailed calculations of phonon frequencies and electron-phonon couplings and their comparison with experiments \cite{cmv-weber}.                                      
\bibliography{at}
\bibliographystyle{unsrt}

 \end{document}